\documentclass[amsmath,twocolumn,secnumarabic,amssymb, nobibnotes, aps, prd]{revtex4-1}
\usepackage{pstricks}
\usepackage{dsfont}
\usepackage{color}
\usepackage{epsfig}
\usepackage{enumitem}
\usepackage{hyperref}

\def\leqn#1{(\ref{#1})}

\begin{document}

\title{Reducing the Quadratic Divergence in the Higgs Mass Squared  Without Top Partners}
\author{Sonia El Hedri$^1$, Ann E.~Nelson$^2$ and Devin G.~E.~Walker$^3$}
\affiliation{$^1$NIKHEF, Theory Group, Science Park 105, 1098 XG, Amsterdam, The Netherlands,}
\affiliation{$^2$Department of Physics, Box 1560, University of Washington, Seattle, WA 98195-1560 USA}
\affiliation{$^3$Department of Physics and Astronomy, Dartmouth College, Hanover, NH 03755 USA}
\begin{abstract}
% Veltman proposed that the Higgs mass should be predicted by the criterion that the quadratic divergence of the Higgs mass cancels.  We now know this is not the case in the minimal standard model.  In this paper, we generalize the Veltman condition to naturally address the little hierarchy problem.  The result is economical class of models that ensures a minimum amount of fine-tuning on the bare Standard Model Higgs mass.  To demonstrate this, we provide a model with an extended Higgs sector and a top Yukawa coupling that is no longer unity.  The latter alleviates the largest radiative corrections to the Higgs mass.  The former features significant dimension-full couplings that generate, at best, logarithmic corrections to the Higgs masses.  The dimensionless couplings are small and do not generate large quadratic divergences or produce Landau poles at energies below the cutoff.  We find the cutoff can be raised to 5 TeV with mild fine-tuning.  We briefly discuss some of the phenomenology of the model and emphasize the importance of future $\bar{t}t h$ measurements.
We examine a model with multiple scalar fields to   see whether it is possible to reduce the fine-tuning of the SM Higgs mass without introducing low scale top partners. Our approach may be regarded as a generalization of   the condition proposed by Veltman, who attempted to predict the Higgs mass using 
the criterion that the various low energy contributions to the quadratic divergence of the Higgs mass cancel.   Although the Veltman condition predicts the wrong Higgs mass in the Standard Model, it can still be adapted to extended Higgs sectors. Furthermore, theories with additional Higgs bosons can lead to suppressed Yukawa couplings of the top quark to the $125$~GeV Higgs, making the associated one-loop divergence smaller. Here, we review   possible extensions of the Standard Model where the Veltman condition could be realized, and study in detail one minimal model with two extra scalar fields. For this model and for a cutoff  of 5~TeV, we show that the overall fine-tuning can be considerably lowered without introducing low-scale Landau poles, albeit the Higgs sector will be strongly coupled at the cutoff. Models where the top Yukawa coupling is reduced, in particular, will be within the reach of the upcoming LHC searches. 
\end{abstract}
\maketitle
 \section{Introduction}
\label{sec:intro}

The discovery of the Higgs boson at the Large Hadron Collider (LHC)~\cite{Aad:2012tfa,Chatrchyan:2012xdj} is a triumph for particle physics, as it was one of the last missing pieces needed to understand the origin of the masses of the Standard Model (SM) particles. Although the SM is now consistent up to high scales, some aspects of the theory still appear  contrived. In particular, the mass of the Higgs boson, 125 GeV, is sensitive to physics at much higher scales. At one-loop, the SM Higgs mass squared  receives sizable  corrections that depend quadratically on the cutoff energy scale, $\Lambda$, as follows
\begin{equation}
m_h^2 = m_0^2  + {1 \over 16 \pi^2} \left( {3 \over 4} g_1^2 + {9 \over 4} g_2^2 + 3 \lambda_h^2 - 12\lambda_t^2 \right)\Lambda^2 \,,
\label{eq:diverge}
\end{equation}
where % we define $\delta m_h^2$ as the quadratically diverging terms in parenthesis.  
$\lambda_h$, $\lambda_t$, v, $g_2$ and $g_1$ are the Higgs quartic, top Yukawa, electroweak vev, $SU(2)_L$ and hypercharge gauge couplings, respectively.  $m_0$ is the bare Higgs mass parameter that appears in the Lagrangian prior to renormalization and $m_h$ is the renormalized quadratic term which determines the value of the physical Higgs mass.  $\Lambda$ is a momentum cutoff, presumably the next fundamental scale of new physics.  In the minimal Standard Model, with a cutoff of $\Lambda \sim 5$~TeV, the value of $m_0$ must be fine-tuned %with the $\Lambda^2$ term 
at the level of one part in 100.  %
This hefty dependence of the physical Higgs mass on the cutoff requires a cancellation between physics at the cutoff scale against physics below the cutoff scale which  is at odds with the expectation that physics at low energies should not be highly sensitive to the short distance theory ~\cite{Wilson:1971bg,Wilson:1973jj}. %Moreover 
Before the discovery of the Higgs boson, accompanied by the lack of discovery of supersymmetry or any other mechanism for cancelling quadratic divergences, it had been expected that  the  scale of new physics beyond the SM would be around or lower than a TeV in order to avoid fine-tuning the SM parameters. The LHC, however, can directly probe  much higher energy scales, as high as several TeV, and has discovered no new physics beyond the SM.   Moreover precision electroweak measurements, which favored a  light SM Higgs boson, imply that $\Lambda$ is likely large ($\Lambda \gtrsim  5$ TeV)~\cite{Barbieri:2000gf,Ciuchini:2014dea} if the new physics particles have electroweak quantum numbers and couple to the SM Higgs.

This gap between the scale of new physics implied by avoidance of fine tuning   and the scale of new physics forecasted by precision electroweak measurements is known as the little hierarchy problem (LHP).  
The traditional way to address the LHP has been to add new particles 
and symmetries to the SM %.
to ensure the one loop quadratic divergence in
%These symmetries ensure the quantum mechanical corrections to the SM Higgs mass, the $\Lambda^2$ term 
equation~\leqn{eq:diverge} %, 
is cancelled~\cite{Schmaltz:2002wx} without large precision electroweak corrections.  %To be consistent with LHC searches, this 
%
% this requires the partner to be decoupled in order to be unseen in LHC searches.  %Because the divergences are largely induced by top quarks, it has been sufficient to just cancel the corrections generated by the top quark partners~\cite{Katz:2005au}.  
Notably, since the largest divergence %in equation~\leqn{eq:diverge} 
is generated by top quarks, extending the SM with fermionic or bosonic top partners is  sufficient to quantitatively solve the LHP~\cite{Katz:2005au,Cohen:1996vb}.  %
%
%relate the SM top quark-Higgs coupling to the top partner-Higgs couplings.  This is done so the 
%allow for quantum mechanical corrections to the SM Higgs mass from SM top quarks loops to be cancelled by the corrections from the top quark partners.  
Introducing such top partners has thus played a large role in theoretical particle physics for the last thirty-five years.  Prominent examples include extra-dimensions~\cite{Hosotani:1983xw,Davies:1987ei,Hatanaka:1998yp,Randall:1999ee}, Intermediate Higgs (rebranded in the literature as natural composite Higgs)~\cite{Katz:2005au}, Little Higgs~\cite{ArkaniHamed:2002qy}, Twin Higgs~\cite{Chacko:2005pe} and Supersymmetry~\cite{Dimopoulos:1981zb,Cohen:1996vb}.  In order for the cancellations to be effective in these models, though, the top partners must have a mass not too far above the top quark mass.  However, the LHC has placed such strong lower bounds on the masses of  colored top partners~\cite{ICHEP:2016ab} that many of the scenarios listed here are now fine-tuned in most of their parameter space.  %
%
% 
%
%Most 

While the colored top partners are currently under siege, an extended Higgs sector is still relatively unconstrained by the LHC.  Although the Higgs couplings to gauge bosons have been measured with high precision, the uncertainties on its couplings to fermions, notably to top quarks, are relatively large~\cite{Aaboud:2017jvq,Sirunyan:2018hoz,Khachatryan:2016vau}. In the light of these results, it is particularly tempting to look for a solution to the fine-tuning problem in either or both of these blind spots.   %
%they are associated with important caveats, especially regarding perturbativity  
%These models however typically consider small deviations of the $125$~GeV Higgs couplings from the SM ones~\textbf{Sonia: is this true?} and thus do not make full use of the current experimental uncertainties in the Higgs sector. 
%
In what follows, we study how to exploit these uncertainties to cancel or at least alleviate the fine-tuning of the Higgs boson mass. 

%Both ATLAS and CMS have measured SM Higgs decays into the gauge bosons as well as $b\bar{b}$ and $\tau\bar{\tau}$~\cite{Khachatryan:2016vau}.  The fact that no significant deviations from the SM have been observed implies %
%
%These results suggest 
%only a minimal mixing between the SM Higgs boson and any new scalars.  
%Because the little hierarchy problem implies no new physics up to a multi-TeV cutoff, one could consider the sum of these experimental results disfavors a naturally light SM Higgs mass.  Instead, 
%
% Additionally, measurements of the $\bar{t}t h$ couplings are still in their relative infancy.  We take the perspective that the results provide a roadmap to write down simple mechanisms %, chosen by nature, 
% which naturally stabilizes the Higgs mass.  Finally, in the absence of the discovery of top quark partners, %an approach that has been gaining popularity in the last fifteen years %has been 
% %is to 
% it has become increasingly popular to simply fine-tune the Higgs mass parameter, $m_0$, relative to the cutoff $\Lambda$ in order to obtain the measured Higgs mass.  This unnatural approach ultimately requires assumptions about the %of $\Lambda$ and 
% %new physics scenario
% %UV completion at the
% UV completion (new physics at cutoff, $\Lambda$). %
% %beyond the SM 
% % ().  
% %To date, there are no known physical mechanisms that can can precisely balance $m_0$ and $\Lambda$ to generate a light mass for Higgs boson.  
% Then the values of the weak scale and the measured SM Higgs mass are simply the result of a coincidental tuning of the value  of $m_0$.

\subsection{Parametric Naturalness}
\label{sec:idea}
Instead of the divergence structure shown in equation~\leqn{eq:diverge}, we consider a scenario where the largest one-loop corrections to the Higgs mass has the following form~\footnote{We note the different sectors in the above equation could conceivably have different cutoffs.  We restrict to equation~\leqn{eq:newdiverge} for simplicity.}%~\footnote{We note each sector may have different cutoffs which will alter and generalize equation~\leqn{eq:newdiverge}.  We do not consider it here and instead focus on a simpler scenario.}
\begin{eqnarray}
m_h^2 &=& m_0^2 \label{eq:newdiverge} \\  
&+&{1 \over 16 \pi^2} \left( {3 \over 4} g_1^2 + {9 \over 4} g_2^2 + 3 \lambda_h^2 - 12\lambda_t^2 + \sum_i c_i \,\lambda_i^2  \right)\Lambda^2 \,.\nonumber
\end{eqnarray}
Here for simplicity we assume a common cutoff scale, $\Lambda$.  The $\lambda_i$ typically correspond to new quartic couplings, present in extended Higgs sectors. Given the cutoff $\Lambda~>~m_h$, there are several strategies to make $m_h$ naturally small:
%\newcounter{saveenum}
%\begin{enumerate}
\begin{itemize}
\item[1.] The $\lambda_i$ summation can be chosen so that the parameters in the last term of equation~\leqn{eq:newdiverge} sum to a small number.  The physical mass, $m_h$, is mostly set by the bare mass, $m_0$.
\end{itemize}
%\setcounter{saveenum}{\value{enumi}}
%\end{enumerate}
This requirement is  known as Veltman's condition~\cite{Veltman:1980mj}.  Veltman, inspired by Wilsonian effective field theories~\cite{Wilson:1971bg,Wilson:1973jj}, insisted the dimensionless parameters in equation~\leqn{eq:diverge} sum to zero.   %After the measurement of the top mass, t
This requirement predicted a 316~GeV SM Higgs mass for the minimal SM, which is not realized in nature.   %As described in the Introduction, the program of adding new symmetries to cancel the one-loop cancellations has dominated theoretical particle physics.  
%Since then, as mentioned in the introduction, the program of adding new symmetries and particles to cancel the one-loop divergences of the SM has dominated theoretical particle physics.  % \leqn{eq:newdiverge}.    %
%
%
In order to balance the top Yukawa coupling contribution in equation~\leqn{eq:newdiverge}, in a renormalizable theory at least some of the new particles running in the loop must be bosonic so that the coefficients $c_i$ are positive. New gauge bosons with electroweak quantum numbers are generically constrained by the LHC to be at least multi-TeV in mass~\cite{Aaboud:2017buh,Aaboud:2017efa,CMS:2017ilm}.  The LHC constraints on new scalars, although weaker, can still push their masses up to almost a TeV.  When the scalar masses are set by their quartic couplings and vevs alone, this result in turn leads to Landau poles well before the LHP can be solved~\cite{Barbieri:2006dq,Barbieri:2005kf}.  In order to ensure a sizable separation between the new and SM Higgs masses, we therefore consider  models involving dimensionful trilinear couplings between the different scalars. Due to gauge invariance, such models will necessary involve not only doublets, but also at least one scalar singlet or triplet that gets a vev.

%%%  In this paper we reconsider how Veltman's condition could be realized in extensions of the SM, without introducing any new symmetry.  We add an economical number of new states in equation~\leqn{eq:newdiverge} with couplings arranged such that the one-loop corrections cancel or nearly cancel at the weak scale.  %

Using extended Higgs sectors to address Veltman's condition has been done in the literature in various ways.  For example, \cite{Masina:2013wja,Karahan:2014ola,Chakrabarty:2018yoy,Darvishi:2017bhf} feature an extended Higgs sector with masses that generate additional quadratic divergences and hierarchy problems.  For many extended Higgs sectors in the literature, it is thus impossible to eliminate all of the quadratic divergences~\cite{Chakrabarty:2018yoy,Kim:2018ecv,Darvishi:2017bhf,Biswas:2016bth,Khan:2017xyh,Chabab:2016vqn,Plascencia:2015xwa,Biswas:2014uba,Kobakhidze:2014afa,Chakraborty:2014oma,Chakraborty:2014xqa,Karahan:2014ola,Antipin:2013exa,Masina:2013wja,Jora:2013opa,Bazzocchi:2012de,Chakraborty:2012rb,Bazzocchi:2012pp,Casas:2004gh,Calmet:2003uj,Kundu:1994bs,Ma:2014cfa,Ma:2001sj}.  %Moreover, to solve the LHP it is necessary to raise the cutoff to at least 5 TeV~\cite{Barbieri:2005kf,Barbieri:2006dq}. The new dimensionless couplings often reach Landau poles well before 5~TeV.  Other work considers new dark matter candidates which mix with the SM~\cite{} as well as
We  introduce two new variations of the Veltman condition, always assuming a single cutoff scale for all loops with a simple momentum cutoff:
\newcounter{saveenum}
\begin{itemize}
 \item[2.] The bare mass, $m_0(\Lambda)$, is evaluated at the cutoff and is zero.    Then the renormalized Higgs mass is set by the radiative correction and is proportional to the new cutoff scale, $\Lambda'$. 
\end{itemize}
In this scenario, it is necessary for the coefficient of $\Lambda^2$ in equation~\leqn{eq:newdiverge} to be small but not zero.  The $16\, \pi^2$ suppression helps tremendously; however, %this coefficient requires 
near but not complete cancellations between fermion and bosonic Higgs couplings are still required.  %We  do not find any parameters which allow this condition  to be satisfied either. 
 This requirement is a mixture between Veltman's condition and an implied aim of Coleman and Weinberg to obtain electroweak symmetry breaking entirely from radiative corrections~\cite{Coleman:1973jx}.  We assume that
at the cutoff, the UV physics gives no  contribution to $m_0$, and then that the $\lambda_i$ are such  that the quantum corrections yield the observed mass of the SM Higgs at the weak scale.   

The last and least restrictive variation of the Veltman condition is simply motivated by avoiding fine-tuned cancellations between physics at different scales.    
\begin{itemize}
\item[3.] Both $m_0$ and the radiative correction are non-zero.  \end{itemize}
This might seem like no constraint at all, but we will search for regions in parameter space where  the cancellations between $m_0$ and the 1-loop corrections to the Higgs mass are not  finely tuned, or at least less fine tuned than in the SM.

\section{A Minimal Model}
\label{sec:model}

We focus on building a  theory that addresses the LHP without adding new symmetries and partners to cancel the one-loop correction shown in equation~\leqn{eq:diverge}.
 %which, to date, are poorly constrained by electroweak precision measurements and the LHC.  %and whose potential to alleviate the fine-tuning of the SM Higgs mass has been underexplored to date. We notably show that models with multiple Higgs bosons can significantly reduce this fine-tuning, allowing to push the cutoff scale $\Lambda$ beyond the reach of the LHC.
%
%In this section, we consider a simple model that embodies the requirements from Section~\ref{sec:idea}. In particular, 
%
In part we are exploiting the fact that the top Yukawa coupling to the 125 GeV Higgs is still not measured precisely at the LHC. The most precise direct measurement  allows this coupling to be reduced by as much as 26\% at the 95\% c.l.~\cite{Aaboud:2017jvq,Sirunyan:2018hoz,Khachatryan:2016vau}. This  can allow  for a reduction of the overall fine-tuning.   It is possible that the top quark receives part of its mass from the vev of a heavier Higgs. We thus consider a two Higgs doublet model. We also 
 add a neutral, real scalar field, $\Phi$, in order to allow for soft trilinear scalar couplings. We consider  the standard Type-II two Higgs doublet model (THDM)~\cite{Gunion:1989we}, although, since the main fermion coupling we are interested in is the top, we expect similar results in other THDM variants.    We could also consider allowing the both doublets to couple to the top quark, which  could introduce flavor changing neutral scalar coupling into the up-quark sector. Since these couplings to light quarks are very small, this type of FCNC is typically compatible with the current experimental constraints.

\subsection{Requirements}  %Parametric Naturalness from an Extended Higgs Sector}

A first step when constructing a mechanism to alleviate the fine-tuning problem is to realize that, in the SM, the corrections to the Higgs mass shown in equation~\leqn{eq:diverge} are largely dominated by the top quark contributions. The size of these contributions, however, strongly depends on the size of the top Yukawa coupling to the Higgs, $\lambda_t$. Although this coupling is   about one in the SM, it has not been precisely measured in the LHC yet. Lower Yukawa couplings would allow to significantly reduce the amount of fine-tuning currently associated to many theories of new physics, whose energy scales have been pushed to a few TeV by the LHC. In what follows, we therefore investigate how simple extensions of the SM could lead to reduced top quark couplings to the Higgs and to what extent these different models would reduce the fine-tuning of the Higgs mass.

The large value of the top Yukawa coupling in the Standard Model is necessary to explain the observed top quark mass of $174$~GeV. In order to reduce the large contribution to the higgs mass radiative correction associated with this coupling, it is therefore necessary to introduce either new vector-like fermions that mix with the top quark, or new Higgs bosons that will provide additional contributions to the top mass. The first approach has already been thoroughly explored by~\cite{Katz:2005au}. Here, we focus on the second scenario and investigate models involving two Higgs doublets  $H_1$ and $H_2$ and a Higgs singlet $\phi$. Although all three Higgses get vevs, only $H_2$ will couple to the top quark. The observed light Higgs is a linear combination of $H_1$ and $H_2$ and so could have a reduced top quark Yukawa coupling, while the Higgs with a larger coupling to the top could be much heavier. The next section details the couplings of this model and the associated constraints from the LHC searches. After EWSB, one of the Higgs mass eigenstates becomes the $125$~GeV Higgs and couples to the top with a strength proportional to the mixing angle between the neutral Higgses.  Flavor changing neutral currents (FCNC) can be completely avoided if the other fermions also only couple to $H_2$, or if leptons and/or the down type quarks couple only to $H_1$. As the couplings of the Higgses to fermions other than the top quark do not lead to meaningful constraints from fine tuning, and can always be made consistent with the FCNC, we   do not study them in detail.

 Before the discovery of the Higgs boson, it was well-known that a heavy SM Higgs boson would help to alleviate the little hierarchy problem.  \cite{Barbieri:2006dq,Barbieri:2005kf} used this fact to generate a heavy SM Higgs boson with a naturally raised cutoff.  Such a scenario, however, required large dimensionless couplings in the scalar potential.  This inevitably led to low scale Landau poles.  %
% that augment the divergences from the top sector.  
% 
%Moreover, the scalar potential would rapidly become non-perturbative and could hit a Landau pole before the little hierarchy problem is satisfied.  
Given this, \cite{Barbieri:2006dq,Barbieri:2005kf} were able to raise the cutoff only to $\Lambda \sim 1.5$ TeV for a SM Higgs mass of about $400-600$~GeV. In a one Higgs model, the only way to raise the Higgs mass is to have a large quartic coupling which leads to a Landau pole at a low scale. In multi-Higgs doublet models the possibilities are more diverse. In order to ameliorate any potential problems with large quadratic corrections from  the scalar sector or low scale Landau poles, we need to limit the size of the dimensionless couplings. For two-Higgs doublet models in which we also do not allow large bare quadratic mass terms, this constraint forbids Higgs masses beyond $\mathcal{O}(100)$~GeV and therefore considerably limits the extent to which the fine-tuning can be reduced. 
In order to allow for large Higgs masses, we introduce an additional Higgs singlet $\phi$ that interacts with the two Higgs doublets $H_1, H_2$ via a trilinear term of the form
\begin{align}
    \mathcal{L}_{\phi h_1 h_2}\sim A_h\,\phi\, H_1\,H_2.
\end{align}
The coupling $A_h$ is now dimensionful and its values are only constrained by perturbative unitarity and vacuum stability~\cite{Schuessler:2007av,Betre:2014fva}.  The contribution to the Higgs mass divergence from these couplings is as best logarithmic
\begin{equation}
\delta m_h^2 \sim {A^2 \over 16\pi^2} \log\left(\Lambda^2/m_h^2 \right)\, ,\end{equation} although the coefficient of the log divergence can be large when $A$ is large.

\subsection{Scalar Sector}

 The fields $H_1$ and $H_2$ and the real scalar singlet $\Phi$ get vevs $v_1$, $v_2$, and $u$ respectively, and have the following structure
\begin{align}
    H_1 &= \begin{pmatrix}
    	G^+ \cos\beta + H^+\sin\beta\\
	\dfrac{1}{\sqrt{2}}\left(v_1 + h_1 + i \left(G^0 \cos\beta + A^0\sin\beta \right)\right)
    \end{pmatrix}\\
    H_2 &= \begin{pmatrix}
    	G^+ \sin\beta - H^+\cos\beta\\
	\dfrac{1}{\sqrt{2}}\left(v_2 + h_2 + i (G^0 \sin\beta - A^0\cos\beta)\right)
    \end{pmatrix}\\
    \Phi &= u + \phi. 
    \label{eq:fields}
\end{align}

Hence, in addition to the Goldstone bosons $G^0, G^\pm$, the theory involves three neutral scalars $(h_1, h_2, \phi)$, one pseudoscalar $A^0$ and one charged scalar $H^\pm$. As in the 2HDM, we introduce a mixing angle $\beta$ such that the vevs of the $SU(2)$ doublets can be rewritten as
\begin{align}
    v_1 &= v\,\cos\beta\quad\quad v_2 = v\,\sin\beta
\end{align}
where $v = 246$~GeV is the electroweak vev.

As for the type-I 2HDM, we require our model to be invariant under a $\mathds{Z}_2$ symmetry.  We choose for the Higgs fields to transform under this symmetry as
\begin{align}
h_1 \to h_1 && h_2 \to - h_2 && \Phi \to -\Phi.
\label{eq:shift2}
\end{align}
EWSB however causes this $\mathds{Z}_2$ symmetry to be spontaneously broken.  This is problematic cosmologically because of the formation of domain walls~\cite{Zeldovich:1974uw,Kibble:1976sj}.  We assume one (or all) of the following:  we break this discrete symmetry softly by small terms.  Alternatively,  
there is a low reheating temperature after inflation~\cite{Vilenkin:1984ib}, below the electroweak scale, so the  temperature is never above the phase transition scale, and the domain walls do not form. Finally, the discrete symmetry could originate from a global $U(1)$ at higher energies.  At the $U(1)$ symmetry breaking scale, cosmic strings form. Then, when the domain walls form at the electroweak scale, they end on loops of the previously formed cosmic strings and the whole string-domain wall network  is no longer  stable and rapidly disappears by radiating scalars~\cite{Vilenkin:1984ib}.
\subsubsection{Overview of the Higgs potential}
The most generic potential consistent with the $\mathds{Z}_2$ symmetry discussed above and minimized around the vevs $v_1, v_2$, and $u$ is
\begin{widetext}
\begin{eqnarray}\label{eq:potential}
V &=& \lambda_1 \left( H_1^\dagger H_1 - \frac{v_1^2}{2} - {A_h \,u\, v_2 \over \lambda_1 v_1} \right)^2 + \lambda_2 \left( H_2^\dagger H_2 - \frac{v_2^2}{2} - {A_h \,u\, v_1 \over \lambda_2 v_2} \right)^2 +  \lambda_3 \left( \Phi^2- u^2 - {A_h \,v_1\, v_2 \over 4\,\lambda_3 u} \right)^2 \label{eq:scalarpotential} \\
&+& \lambda_4 \left(  H_1^\dagger H_1 - \frac{v_1^2}{2} +  H_2^\dagger H_2 - \frac{v_2^2}{2}  \right)^2 + \lambda_5 \biggl( H_1^\dagger H_1 \,H_2^\dagger H_2 - H_1^\dagger H_2 \,H_2^\dagger H_1 
%%%
%- {\lambda_1 \over \lambda_5}\left({A_h \,u\, v_2 \over 2 \,\lambda_1 v_1}\right)^2 
%- {\lambda_2 \over \lambda_5}\left({A_h \,u\, v_1 \over 2 \,\lambda_2 v_2}\right)^2 
%-  {\lambda_3 \over \lambda_5}\left({A_h \, v_1\,v_2 \over 4 \,\lambda_3 u}\right)^2 \,\,
\biggr) %\nonumber \\ %+ \lambda_5 \left( \mathrm{Re}\left(h_1^\dagger h_2 \right) - v_1 v_2 \cos\xi \right)^2 
%&+&\lambda_6 \left( \mathrm{Im}\left(h_1^\dagger h_2 \right) - v_1 v_2 \sin\xi \right)^2 \nonumber 
+ \lambda_6 \left( H_1^\dagger H_1 - \frac{v_1^2}{2}  \right)\left( \Phi^2- u^2 \right) \nonumber \\
&+&  \lambda_7 \left( H_2^\dagger H_2 - \frac{v_2^2}{2} \right)\left( \Phi^2- u^2 \right) + A_h \left( \Phi\, H_1 H_2^\dagger + \Phi \,H_1^\dagger H_2  - u\, v_1 v_2\,\cos \xi \right). \nonumber
\end{eqnarray}
\end{widetext}
  %In order not to generate large corrections to the SM Higgs mass from the potential we require $0 \leq \lambda_i < 1$. 
  The last trilinear term in equation~\leqn{eq:potential} generates off-diagonal contributions to the scalar mass matrix of the form $A_h\, u \,h_1 h_2$ and $A_h v_{1,2} \,h_{2,1}\phi$. Minimizing the scalar potential around the vevs also causes this term to contribute to the diagonal elements of this scalar mass matrix as well as to the masses of the pseudoscalar $A^0$ and charged Higgs $H^\pm$. The squared mass matrix for the neutral scalars can then be written as
  \begin{widetext}
  \begin{eqnarray}\label{eq:massmatrix}
      M_h^2 &= \begin{pmatrix}
      	2 v^2 (\lambda_1 + \lambda_4)\cos^2\beta - A_h u\tan\beta & A_h u + v^2 \lambda_4\sin(2\beta) & v(2 u \lambda_6 \cos\beta + A_h\sin\beta) \\
	A_h u + v^2 \lambda_4\sin(2\beta) & -A_h u \cot\beta + 2 v^2 (\lambda_2 + \lambda_4)\sin^2\beta &  v (A_h\cos\beta +2 u\lambda_7\sin\beta)\\
	v(2u\lambda_6\cos\beta + A_h\sin\beta) & v(A_h\cos\beta + 2 u\lambda_7\sin\beta) & 8 u^2\lambda_3 - \frac{A_h}{2u} v^2 \sin(2\beta)
      \end{pmatrix}.
  \end{eqnarray}
  \end{widetext}
Diagonalizing this matrix will give three scalar mass eigenstates $(h, h', h'')$ with masses $m_h$, $m_{h'}$, and $m_{h''}$. As a convention for the rest of this work, we define $h$, $h'$, and $h''$ as the states with the largest $h_1$, $h_2$, and $\phi$ component respectively. In the limit where $\lambda_i \ll 1$ and $|A_h|, v_1, v_2 \ll u$, the mass eigenstates can be approximated by 
\begin{widetext}
\begin{eqnarray}
m_1^2 &=& 2 A_h\, v_1 +\left( 2\left(\lambda_3 \,u^2+\lambda_7 \,u\, v_2+ (\lambda_2+\lambda_4)\,v_2^2 \,\right)- \frac{ \left( \,\left(2 \lambda_3-2 \lambda_6+\lambda_7 \right) u + (2 \lambda_2-2 \lambda_4+\lambda_7) v_2\right)(u+v_2)v_1^2}{u\, v_2}\right) \hspace{0.5cm}\nonumber \\
&+& \frac{1}{2 A_h v_1 v_2^2}\biggl( 2 \lambda_3 \left(\lambda_3 u^2 -(\lambda_2+\lambda_4)v_2^2\right) u^2 v_2^2 \\
&+& \left(\lambda_3^2\, u^4+\lambda_3 (2 \lambda_2-3 \lambda_3-6 \lambda_4+4 \lambda_6) \,u^2 v_2^2 - (\lambda_2+\lambda_4) (3 \lambda_2-2 \lambda_3-5 \lambda_4+4 \lambda_6) \,v_2^4 \right)v_1^2  \biggr)+ \ldots \nonumber \\
m_2^2 &=& -2 A_h v_1 +  \left( 2 \lambda_3\, u^2 - 2 \lambda_7 \,u \,v_2+2 (\lambda_2+\lambda_4)\,v_2^2 +\frac{  \left( (2 \lambda_3-2 \lambda_6+\lambda_7)\,u - (2 \lambda_2-2 \lambda_4+\lambda_7)\,v_2 \right)(u-v_2)\,v_1^2}{u \,v_2} \right)\nonumber  \\
&-& \frac{1}{2 A_h v_1 v_2^2}\biggl(2 \lambda_3 u^2 v_2^2 \left(\lambda_3 u^2-2 (\lambda_2+\lambda_4)v_2^2\right)+ \bigl(\lambda_3^2 \,u^4+\lambda_3 (2 \lambda_2-3 \lambda_3-6 \lambda_4+4 \lambda_6)\, u^2 v_2^2 \\
&-&  (\lambda_2+\lambda_4) (3 \lambda_2-2 \lambda_3-5 \lambda_4+4 \lambda_6) v_2^4 \bigr) v_1^2 \bigr) + \ldots \nonumber \\
m_3^2 &=& 4 \left(\lambda_1+\lambda_2+\lambda_3-\lambda_6+\lambda_7 \right) v_1^2 -A_h \left(\frac{v_1 \,v_2}{u} + \frac{u \,v_1}{v_2} + \frac{u \,v_2}{v_1}\right) + \ldots 
\end{eqnarray}
\end{widetext}
 In Appendix~\ref{sec:eig}, we give the neutral mass eigenstates for $\lambda_i \ll 1$ and $v_1, v_2, u \ll |A_h|$.  Throughout this work, the scalar field with the largest $h_1$ component $h$ is taken to be the observed Higgs boson, with $m_h = 125$~GeV. In order for $h'$ and $h''$ to be heavier than $h$, $A_h$ should be negative. Thus, the addition of the singlet $\phi$ and its associated trilinear coupling to the model helps to establish a significant mass hierarchy between the $125$~GeV Higgs and the extra scalar fields without having large quartic couplings. The mass eigenstates $(h, h', h'')$ are related to the interaction eigenstates $(h_1, h_2, \phi)$ by the following rotation matrix
\begin{align}
    \label{eq:rotmatrix}
    \begin{pmatrix}
    	h\\
	h'\\
	h''
    \end{pmatrix} &=
    \begin{pmatrix}
    	c_1 c_3 - c_2 s_1 s_3 & c_3 s_1 + c_2 c_1 s_3 & s_2 s_3\\
	-c_1 s_3 - c_2 c_3 s_1 & c_1 c_2 c_3 - s_1 s_3 & c_3 s_2\\
	s_1 s_2 & -c_1 s_2 & c_2
    \end{pmatrix}\begin{pmatrix}
    	h_1\\
	h_2\\
	\phi
    \end{pmatrix}
\end{align}
where the $c_i$, $s_i$ represent the cosines and the sines of the Euler angles $\alpha_i$. We can see that the $\alpha_2$ angle drives the mixing of $\phi$ with the $SU(2)$ doublet states. When this mixing is small, the angle $\alpha = \alpha_1 + \alpha_3$ can be interpreted as the mixing angle between $h_1$ and $h_2$. In this limit, our model becomes similar to a two-Higgs doublet model. As we will see in the rest of this section, our main difference with a standard  2HDM will be that in our scenario, the up-type quarks will couple preferentially to the BSM Higgs bosons in the limit where $\alpha$ is low.

The masses of the charged ($H^{\pm}$) and CP-odd neutral ($A^0$) Higgs bosons can be obtained analytically as follows
\begin{eqnarray}
m_{H^{\pm}}^2 &=&  \frac{\lambda_5}{2} v^2 -\frac{A_h u}{\cos\beta\sin\beta}  \\
m_{A^0}^2 &=& -{A_h \,u \over \cos\beta\sin\beta}.
\label{eq:mAH}
\end{eqnarray}
Here, the role of the trilinear term $A_h\,h_1\,h_2\,\phi$ in generating a mass hierarchy between the SM Higgs $h$ and the BSM Higgses is obvious as this term leads to the $|A_h|u$ contributions in equation~\eqref{eq:mAH}.
%In some parts of this work, in order to acquire a better intuition of how the couplings of the Higgs mass eigenstates change with respect to the SM, we use the fact that 
\newline
\subsubsection{Vacuum stability and perturbativity}
In order for the Higgs potential shown in equation~\leqn{eq:potential} to be valid, it needs to be bounded from below and the quartic and trilinear couplings need to satisfy perturbativity and unitarity requirements. In order for the potential to not go to minus infinity for large values of the scalar fields, we require the quartic couplings satisfy the following conditions, taken from~\cite{Drozd:2014yla}
\begin{widetext}
\begin{align}
&\lambda_{1,2} + \lambda_4 > 0 && |\lambda_{6}| < 4\sqrt{\lambda_3 (\lambda_1 + \lambda_4)}\\
&\lambda_3 > 0 &&  |\lambda_{7}| < 4\sqrt{\lambda_3 (\lambda_2 + \lambda_4)} \\
& \lambda_4  + \lambda_5> -\sqrt{(\lambda_1 + \lambda_4)(\lambda_2 + \lambda_4)} &&\lambda_4 + \lambda_5/2 > -\sqrt{(\lambda_1 + \lambda_4)(\lambda_2 + \lambda_4)} .
\end{align}
In addition for $\lambda_6 < 0$ or $\lambda_7 < 0$, we also require
  \begin{align}
  %    &\lambda_1 + \lambda_4 > 0\nonumber\\ &\lambda_2 + \lambda_4 > 0\nonumber \\ &\lambda_3 > 0\nonumber\\
%      &\lambda_4  + \lambda_5> -\sqrt{(\lambda_1 + \lambda_4)(\lambda_2 + \lambda_4)}\nonumber\\ &\lambda_4 + \lambda_5/2 > -\sqrt{(\lambda_1 + \lambda_4)(\lambda_2 + \lambda_4)}\\
	%&|\lambda_6| < 4\sqrt{\lambda_3 (\lambda_1 + \lambda_4)}\nonumber\\& |\lambda_7| < 4\sqrt{\lambda_3 (\lambda_2 + \lambda_4)}\nonumber\\
	 -\frac{1}{2}\lambda_6\lambda_7 + 4\lambda_3 (2\lambda_4 + \lambda_5) &> -\sqrt{4\,(4 \lambda_3 (\lambda_1 + \lambda_4) - \lambda_6^2/4)(4 \lambda_3 (\lambda_2 + \lambda_4) - \lambda_7^2/4)} \\
	-\frac{1}{2}\lambda_6\lambda_7 + 8\lambda_3 (\lambda_4 + \lambda_5) &> -\sqrt{4\,(4 \lambda_3 (\lambda_1 + \lambda_4) - \lambda_6^2/4)(4 \lambda_3 (\lambda_2 + \lambda_4) - \lambda_7^2/4)}.
  \end{align}
  \end{widetext}

In addition to the above requirements, the quartic \mbox{couplings} $\lambda_i$ also need to remain perturbative up to at least the cutoff scale $\Lambda$ at which new physics should appear. In the rest of this work, for each scale $\Lambda$ that we consider, we require the scalar quartic couplings to satisfy $|\lambda_i| < 4\pi$ and still fulfil the vacuum stability conditions at the cutoff. Our RGEs for $\lambda_i$, $A_h$, and the top quark Yukawa coupling are shown in Appendix C. \ Besides an extended Higgs sector, our model will have modified couplings of the $125$~GeV Higgs to fermions and gauge bosons. While the couplings of the Higgs to gauge bosons have been measured to be within $10$\% of the SM ones, the measurements of the Higgs couplings to fermions have either been indirect or with large uncertainties. In the rest of this section, we discuss the impact of the Higgs-gauge coupling measurements on the mixing angles of the neutral scalars and how the large uncertainty on the Higgs to top coupling measurement can be exploited to alleviate the fine-tuning problem.

\subsection{Yukawa couplings to fermions}
\label{sec:yukawa}

%In this paper, we primarily focus on the couplings between the top quark and the Higgses.  In order to forbid flavor changing neutral currents, the top quark couplings described below are applied to all up-type quarks.
%

\begin{figure}
    \includegraphics[width=\columnwidth]{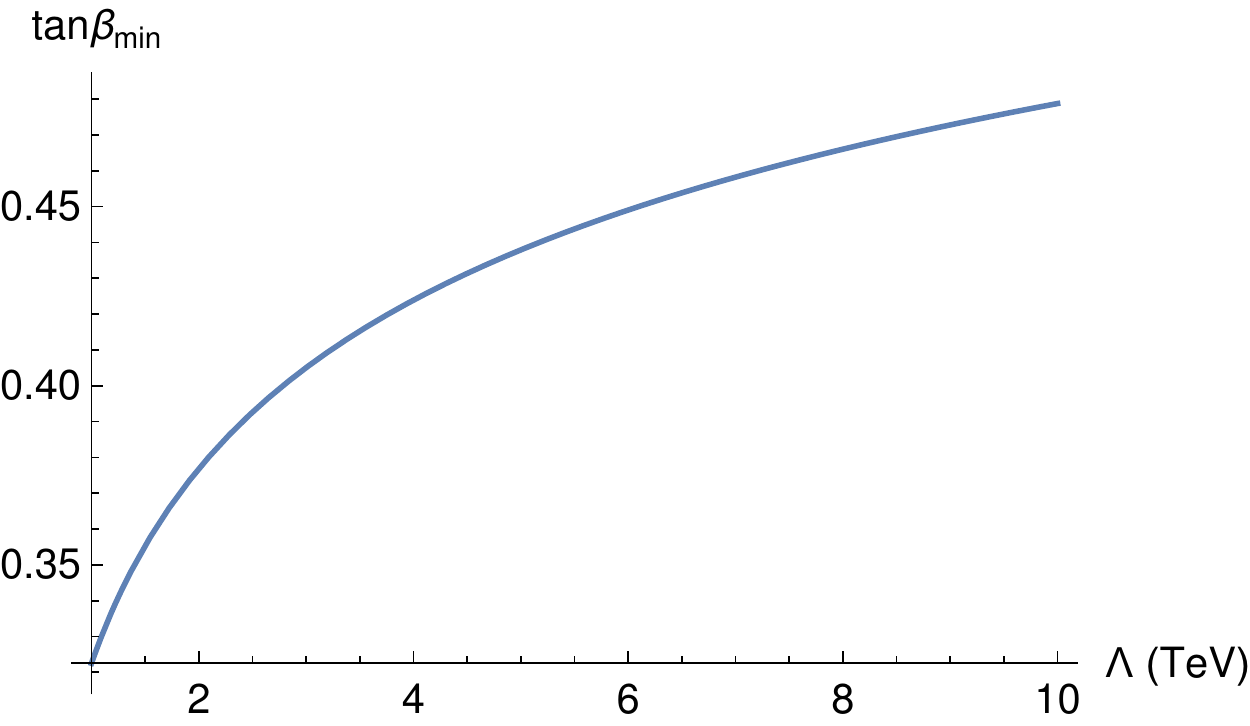}
    \caption{\label{fig:ypert} Minimum value of $\tan\beta$ allowed by requiring no Landau poles below $\Lambda$ for the top quark coupling to $h'$.}
\end{figure}

In standard two-higgs doublet models, the up-type quarks couple preferentially to  the interaction eigenstate that contributes mostly to the SM-like Higgs $h$ in the low mixing limit. In our model, in order to obtained a suppressed coupling of the $125$~GeV Higgs to the top quark, we assume that $h$ is mostly $H_1$ while   the up-type quarks   couple only to $H_2$.  This constraints can be the result of a $\mathds{Z}_2$ symmetry under which the right handed up-type quarks such as $t_R$ transform  as  
\begin{equation}
t_R \to - t_R
\end{equation}
and $H_2\to -H_2$. In order to avoid problematic FCNCs, we may take the down-type quarks  and leptons to couple only to either $H_1$ or to $H_2$.  The couplings of the Higgs bosons to the quarks can then be written as
\begin{equation}
\mathcal{L}_t = \lambda_u \,q_L\, u_R \,h_2 + \lambda_d \, q_L\, d_R\, h_1.
\label{eq:lambdaud}
\end{equation}
The couplings of the Higgses to leptons will not be relevant for this study. In the rest of this work, we will focus particularly on the couplings of the Higgses to the top and the bottom quarks, that can be written as
\begin{equation}
\mathcal{L}_t = \lambda \,q_L\, t_R \,h_2 + \lambda_b \, q_L\, b_R\, h_1.
\label{eq:lambda}
\end{equation}
% In terms of the mass eigenstates, the top Yukawa coupling to the 125 GeV Higgs is 
% \begin{widetext}
% \begin{eqnarray}
% \lambda_t &=& \frac{A_h u v_2^2}{2 v_1^2 (\lambda_1+\lambda_4) \left(u^2+v_2^2\right)} -\frac{2 u^3 v_2^2 \left(u^2 (2 \lambda_1+\lambda_6)+v_2^2 (-2 \lambda_1-4 \lambda_4+\lambda_6)\right)}{A_h \left(u^2+v_2^2\right)^3} + \frac{u^4 v_2 (2 (\lambda_1+\lambda_4)+\lambda_6)+u^2 v_2^3 (\lambda_6-2 \lambda_1)+2 \lambda_4 v_2^5}{2 v_1 (\lambda_1+\lambda_4) \left(u^2+v_2^2\right)^2}.
% \end{eqnarray}
% \end{widetext}
 In order to obtain the correct top and bottom quark masses $m_t$ and $m_b$, the strength of the couplings $\lambda$ and $\lambda_b$ should be the following
\begin{align}
    \lambda &= \frac{\lambda_t^\mathrm{SM}}{\sin\beta}\quad\quad\lambda_b = \frac{\lambda_b^\mathrm{SM}}{\cos\beta}.
\end{align}
The coupling of the top quark to $h'$ is thus typically larger than $1$, which could lead to low scale Landau poles. Requiring no Landau poles for $\lambda_t'$ up to a given cutoff scale $\Lambda$ leads to an upper bound on the value of $\lambda_t'$ at the EW scale, that translates in turn into a lower bound on $\beta$. This lower bound is shown as a function of $\Lambda$ in figure~\ref{fig:ypert}. The RGEs for the top quark and strong couplings that we solved to obtain this result are shown in Appendix~\ref{sec:RGE}. Conversely, the LHC measurements of the $125$~GeV Higgs coupling to $b\bar{b}$ as well as perturbativity requirements on $\lambda_b$ should provide an upper bound on $\beta$. Since the uncertainties on the Higgs coupling to b-quarks measurements at the LHC are huge and $\lambda_b^\mathrm{SM}$ is very small, however, this upper bound is expected to be extremely loose and we do not take it into account in the rest of this work. 

When the mixing angles between the scalars are small, the top quark should couple preferentially to $h'$ ---the mass eigenstate most similar to $h_2$--- while the top coupling to $h$
, $\lambda_t$, will be mixing suppressed. This coupling can be written as a function of the mixing angles in equation~\eqref{eq:rotmatrix} to obtain
\begin{align}
    \label{eq:lambdatrot}
    \lambda_t &= \lambda_t^\mathrm{SM} \frac{c_3 s_1 + c_2 c_1 s_3}{s_\beta}.
\end{align}
In the limit where the $\phi$ mixing to $h_1$ and $h_2$ is small, this coupling can be approximated by 
\begin{align}
    \label{eq:lambdatrot2}
    \lambda_t &\approx \lambda_t^\mathrm{SM} \frac{\sin\alpha}{\sin\beta}
\end{align}
where $\alpha = \alpha_1 + \alpha_3$ is the mixing angle between $h_1$ and $h_2$. In order for this coupling to be significantly lower than the SM coupling, it is therefore crucial to depart from the alignment limit, with $\alpha < \beta$. Going away from the alignment limit could however significantly modify the couplings of the $125$~GeV Higgs to other SM particles. In most cases, these deviations are not  directly highly constrained at the LHC since the Higgs couplings to the other SM fermions have not been precisely measured to date.  The Higgs couplings to photons and gluons do indirectly constrain the fermion couplings more precisely, however these can also be easily modified by introducing higher dimensional operators or other new physics so we do not necessarily need to consider these indirect constraints on the fermion couplings. The Higgs couplings to vector bosons, on the other hand, are tree-level and have been measured to a fairly good level of precision at the LHC. It is therefore crucial to determine how a modification of $\lambda_t$ would affect these couplings in our model.

\subsection{Gauge Sector}
\label{sec:gauge}
\begin{figure}
    \includegraphics[width=\columnwidth]{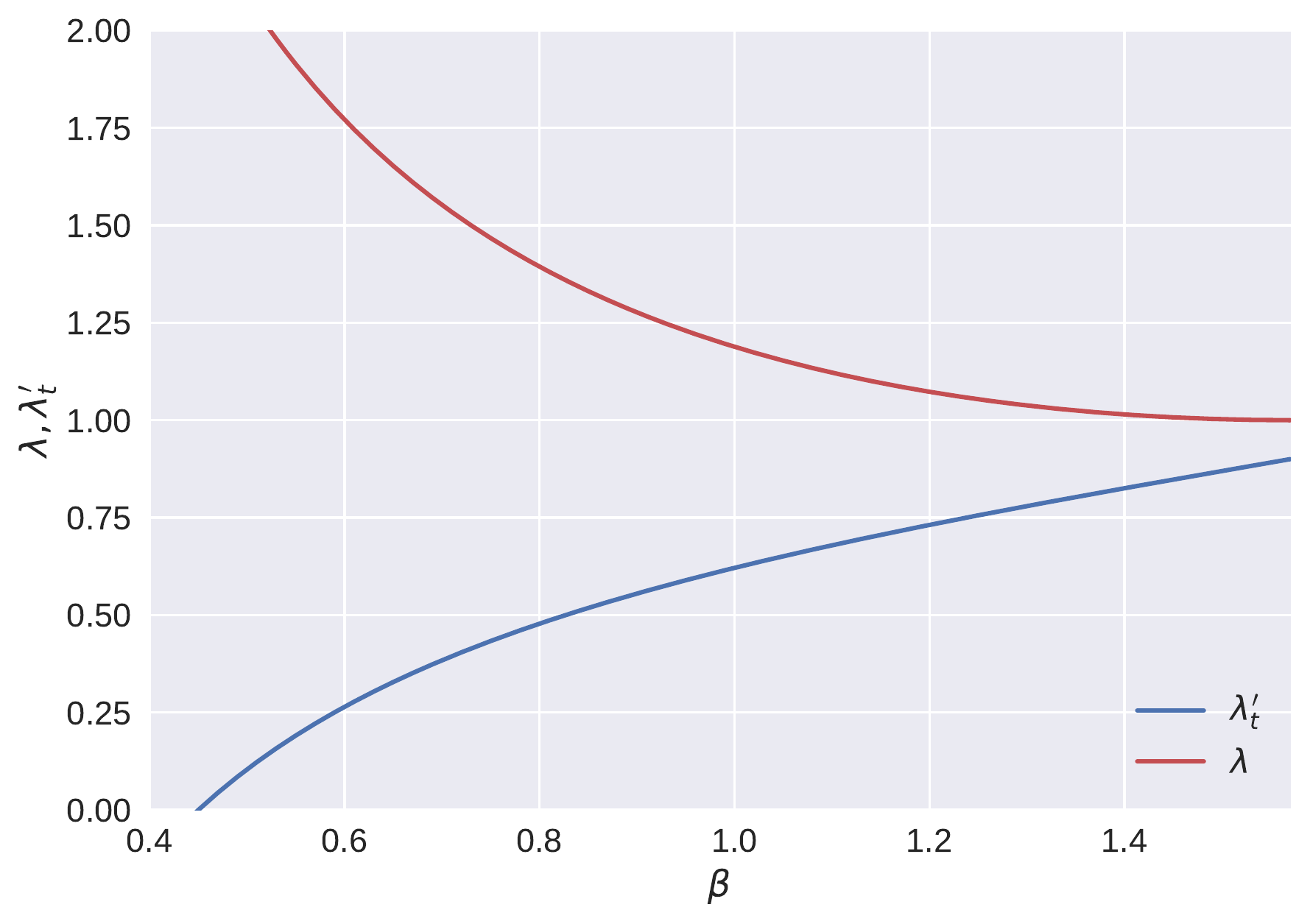}
    \caption{\label{fig:lambdat} Value of $\lambda_t'$ when the mixing angle $\alpha$ between the two Higgs doublets is set to $\alpha = \beta - 0.45$, that is, the minimum value allowed by the Higgs-gauge coupling measurements (blue line). The red line shows the top coupling to $H_2$ $\lambda$, defined in equation~\leqn{eq:lambda}. We can see that, if the mixing angle $\alpha$ is free to vary as far from $\beta$ as allowed by the LHC measurements, it is possible to get an extremely suppressed top coupling to the $125$~GeV Higgs while still having $\lambda$ remain perturbative up to at least $5$~TeV.}
\end{figure}

The electroweak gauge bosons couple to the Higgs doublets through the standard covariant derivative
\begin{equation}
D \mathcal{H} = \partial \mathcal{H} - i g_2 W \mathcal{H} - i g_1 Y \mathcal{H} \, .
\end{equation}
Here $\mathcal{H} = (h_1, h_2)$ and $g_2$ and $g_1$ are the $SU(2)_L$ and $U(1)_Y$ gauge couplings. Using the definitions introduced in equation~\leqn{eq:fields} for the Higgs fields, we can readily check that we obtain the correct masses for the gauge bosons. 
Using the notation defined in equation~\eqref{eq:rotmatrix} for the mixing angles, the light Higgs coupling to the gauge bosons reads as
\begin{align}
	g_{hVV} &= g_{hVV}^{\mathrm{SM}} [c_3\,\cos(\beta - \alpha_1) + c_2 s_3\,\sin(\beta - \alpha_1)]
\end{align}
which is always smaller than one. In order to be consistent with the current LHC results~\cite{Khachatryan:2016vau}, $g_{hVV}$ needs to be of at least $90\%$ of the SM value. In order to understand the implications of this constraints, we consider first the case where $c_2 \approx 1$, which corresponds to a scenario where $\phi$ mixes very little with the scalar components of the Higgs doublets. In this limit, the coupling between $h$ and the gauge bosons becomes
\begin{align}
	g_{hVV} &= g_{hVV}^{\mathrm{SM}} \cos(\beta - \alpha)
	\label{eq:gaugecoupling}
\end{align}
where again $\alpha = \alpha_1 + \alpha_3$. In order for this coupling to be close to the SM value, we therefore need to be in the alignement limit where $\alpha$ and $\beta$ are close to each other. This requirement might be in tension with our end goal of reducing the top coupling to $h$ defined in equation~\leqn{eq:lambdatrot}, the latter being close to one in this limit. The current LHC results however still leave some significant freedom since the requirement that $g_{hVV} \geq 0.9\, g_{hVV}^\mathrm{SM}$ translates into
\begin{align}
	|\beta - \alpha| \leq 0.45.
\end{align}
The values of $\lambda_t$ and $\lambda_t'$ for $\alpha = \beta - 0.45$ are shown as a function of $\beta$ in figure~\ref{fig:lambdat}. As can be inferred from both this figure and figure~\ref{fig:ypert}, it is a priori possible to considerably reduce the value of the top coupling to the $125$~GeV Higgs without significantly reducing its coupling to gauge bosons or introducing Landau poles below at least $5$~TeV from the running of $\lambda_t'$. Note that the top Yukawa coupling is most suppressed for low values of $\beta$, so the coupling of the $125$~GeV Higgs to bottom quarks is SM-like and well within the LHC limits. 

When the singlet $\phi$ mixes with $h_1$ and $h_2$, the reasoning outlined above still applies. Since the mixing between the scalar singlet and the scalar components of the doublets is governed by $\alpha_2$, increasing this mixing would translate into decreasing $c_2$. In order for the $g_{hVV}$ coupling in equation~\leqn{eq:gaugecoupling} to remain close to the SM value, $c_3\cos(\beta - \alpha_1)$ needs to increase. This constraints pushes us into a region of parameter space where $\alpha_3$ is small and the mixing between $h_1$ and $h_2$ is governed by $\alpha_1$. This scenario is qualitatively similar to the one where the singlet $\phi$ is decoupled and the steps detailed above can be repeated with $\alpha_1$ substituted to $\alpha$.

As our results show, strongly reducing the $125$~GeV Higgs coupling to the top quark without creating tension with the current LHC results or introducing low scale Landau poles can be achieved for certain values of the mixing angles between the neutral scalars. It is crucial in particular that the effective mixing angle between $h_1$ and $h_2$ is as far from $\beta$ as allowed by the current Higgs-gauge coupling measurements. This requirement, however, constrains the product $\sqrt{|A_h| u}$, appearing in the mass matrix~\leqn{eq:massmatrix} to not be too large compared to the EW scale. Consequently, the regions of parameter space where the $125$~GeV Higgs coupling to the top quarks is the lowest are also regions where the other Higgses have $\mathcal{O}(100)$~GeV masses. These other Higgses, however, can exhibit large quadratic divergences that could be reduced only by reintroducing sizable fine-tuning. In particular, by construction, $h'$, $A^0$, and $H^\pm$ are associated with order one top Yukawa couplings and therefore need to be much heavier than $h$. In what follows, we discuss how we impose the naturalness requirement and compute the dominant fine-tuning factors.

\section{Naturalness}
\label{sec:natural}
At one loop the quadratic divergences generated by the minimal model for the masses of any of the Higgses can be represented as
\begin{equation}
\delta m_{h_i}^2 = \alpha_t \,\Lambda^2_t + \alpha_g\,\Lambda^2_g + \alpha_h \Lambda_h^2
\end{equation}
where we have followed the notation in~\cite{Barbieri:2006dq}. In what follows, we will neglect the contributions from the gauge boson loops $\alpha_g$ due to the low values of the weak couplings. The quadratic divergences from the top and Higgs boson loops in our model can then be derived from the Coleman-Weinberg potential
\begin{eqnarray}\label{eq:quadraticCW}
V_\mathrm{quadratic} &=& %{\Lambda^2 \over 32\,\pi^2}  \mathrm{tr}\biggl[M_\mathrm{s}^\dagger M_\mathrm{s}  \biggr]%-  {3 \Lambda^2 \over 8\pi^2} M_\mathrm{top}^\dagger M_\mathrm{top}
% \nonumber \\
%&=& 
 {\Lambda^2 \over 32\,\pi^2} \left( \lambda_1+2 \lambda_4+\frac{1}{2}\lambda_5 +\frac{1}{2}\lambda_6\right) H_1^\dagger H_1 \nonumber\\
&+&{\Lambda^2 \over 32\,\pi^2} \left( \lambda_2+2 \lambda_4+\frac{1}{2}\lambda_5+\frac{1}{2}\lambda_7 \right) H_2^\dagger H_2 \\ 
&+& {\Lambda^2 \over 32\,\pi^2} \left(\lambda_3+{\lambda_6 \over 2}+\frac{\lambda_7}{2}\right) \Phi ^2 - {3\,\lambda^2\,\Lambda^2 \over 8\,\pi^2} H_2^\dagger H_2.  \nonumber
%&-&  {3 \Lambda^2 \over 8\pi^2}  \nonumber
\end{eqnarray}
We discuss our derivation in more details in appendix~\ref{sec:CWpotential}. The values of the fine-tuning factors $\alpha_i$ can be deduced by rotating into the mass basis and computing the derivatives of $V_\mathrm{quadratic}$ with respect to the different fields. For the light Higgs $h$, we obtain
\begin{eqnarray}
\alpha_{th} &=& {3 \lambda_t^2 \over 4 \pi^2} \\
    \alpha_{hh} &=& \alpha_{h11} (c_1 c_3 - c_2 s_1 s_3)^2 + \alpha_{h22} (s_1 c_3 + c_2 c_1 s_3)^2 \nonumber\\
    && + \alpha_{h33} (s_2s_3)^2\nonumber
\end{eqnarray}
where $\lambda_t$ is defined in equation~\leqn{eq:lambdatrot} and the $\alpha_{hii}$ are defined by
\begin{align}
    \alpha_{h11} &= -{(2\lambda_1 + 4\lambda_4 + \lambda_5 + \lambda_6)\over 16 \pi^2 }\\
    \alpha_{h22} &= -{(2\lambda_2 + 4\lambda_4 + \lambda_5 + \lambda_7)\over 16 \pi^2 }\\
    \alpha_{h33} &= -{(2\lambda_3 + \lambda_6 + \lambda_7)\over 16 \pi^2 }
\end{align}
and are weighted by combinations of the cosines and sines of the mixing angles $\alpha_{i}$, defined in equation~\eqref{eq:rotmatrix}. The coupling $\lambda_t$ as well as the Higgs quartic couplings are evaluated at the cutoff scale $\Lambda$. Usually, when the quartic couplings $\lambda_i$ are taken to be perturbative, these quadratic divergences lead to a lower fine-tuning than the ones from the top loops. Similarly, we derive the fine-tuning factors for the other two Higgses:
\begin{eqnarray}
\alpha_{th'} &=& {3 \lambda_t'^2 \over 4 \pi^2}\nonumber \\
    \alpha_{hh'} &=& \alpha_{h11} (-c_1 s_3 - c_2 s_1 c_3)^2 + \alpha_{h22} (c_1 c_2 c_3 - s_1 s_3)^2 \nonumber\\
    && + \alpha_{h33} (s_2c_3)^2\nonumber\\
\alpha_{th''} &=& {3 \lambda^2 (c_1 s_2)^2 \over 4 \pi^2} \\
    \alpha_{hh''} &=& \alpha_{h11} (s_1s_2)^2 + \alpha_{h22} (c_1s_2)^2 + \alpha_{h33} c_2^2\nonumber.
\end{eqnarray}
Finally, the pseudoscalar and charged Higgses are also associated with large quadratic divergences. The corresponding fine-tuning factors are
\begin{eqnarray}
\alpha_{t\{A,H^\pm\}} &=& {3 \lambda^2\cos^2\beta \over 4 \pi^2} \\
    \alpha_{h\{A,H^\pm\}} &=& \alpha_{h11} \sin^2\beta + \alpha_{h22} \cos^2\beta \nonumber.
\end{eqnarray}
The sensitivity of the SM Higgs masses to a given cutoff scale $\Lambda_i$ is given by the formula
\begin{align}
D(m_h) = \left|{\partial \log m_h^2 \over \partial \log \Lambda^2_i} \right| = {|\alpha_i| \, \Lambda^2_i \over m_h^2}
\end{align}
In the rest of this work, the fine-tuning factor that we consider at a given scale will be the maximal value of the fine-tunings associated to the top, gauge boson or Higgs loops, for all the three Higgs bosons
\begin{align}
D_\mathrm{max}(\Lambda) = \mathrm{max}_{j=\{h,h',h'',A,H^\pm\}} \left\{|\alpha_{tj} + \alpha_{hj}| \, \Lambda^2 \over m_j^2\right\}.
\label{eq:finetuning}
\end{align}
This estimate is conservative since it assumes that the cutoff scales for all the loop contributions to the Higgs masses will all be at their lowest possible values for a given $D_\mathrm{max}$. Since $\alpha_{tj}$ and $\alpha_{hj}$ are of opposite signs, they are expected to cancel out, either partially, or, if the Veltman condition 1 is fulfilled, totally. 

Besides looking for parameters with low total fine-tuning, we also must consider the current LHC searches for new bosons. In the next section, we detail what searches and decay channels are relevant to our models and how we implement the corresponding constraints.

\section{LHC phenomenology}
\label{sec:lhc}

As highlighted in section~\ref{sec:gauge}, in order to be as far as possible from the alignment limit, it is necessary for either $h'$ or $h''$ to be light, with masses typically below a TeV. If $h'$ is light, in particular, the model will also involve a light pseudoscalar $A^0$ and charged Higgs $H^\pm$, which could both be within the reach of the corresponding LHC searches. At low $\tan\beta$ in the MSSM, which is also the preferred region for our low fine-tuning models as discussed in section~\ref{sec:yukawa}, pseudoscalar Higgses are excluded up to about $400$~GeV. It is therefore crucial to investigate how the different Higgs searches at $13$~TeV LHC will constrain our models.  

As discussed in details in section~\ref{sec:gauge}, the couplings of the $125$~GeV Higgs to the gauge bosons are constrained to be SM-like and the couplings to the bottom quark are expected to be much smaller than one. The couplings of these particles to the new Higgs bosons will therefore be suppressed and the corresponding LHC searches should not be particularly sensitive to our model. Similarly, the top quark couplings to $h''$ is expected to be mixing suppressed, which would lead to reduced gluon fusion production rates. The second Higgs $h'$, as well as $A^0$ and $H^\pm$, however, have the following couplings to the top quarks
\begin{align}
    \lambda'_t \approx \frac{\lambda_t^\mathrm{SM}}{\sin\beta}\quad\quad \lambda_{t}^{A^0, H^\pm} = \frac{\lambda_t^\mathrm{SM}}{\tan\beta}.
\end{align}
When $\tan\beta$ is of order one or lower these couplings will be of same order as, if not larger than, the SM top Yukawa couplings. These new Higgses will therefore have sizable production rates through gluon fusion at the LHC and should therefore be severely constrained by the current searches. 

In MSSM models with $\tan\beta\lesssim 3$, heavy pseudoscalar and charged Higgs bosons have been already excluded up to about $350$~GeV~\cite{Aaboud:2017gsl,Aaboud:2017rel,Aaboud:2017cxo,CMS:2017vpy,CMS:2016ncz}. In order to account for the possible mild suppressions of the production rates of these particles in our model, we focus on parameter points where $h'$, $A^0$, and $H^\pm$ all have masses larger than $250$~GeV, which corresponds to the lowest masses explored by the $13$~TeV LHC Higgs searches. For these masses, the main decay modes are
\begin{align}
    \label{eq:decay}
    h', h'' &\rightarrow t\bar{t}, b\bar{b}, ZZ, hh, W^+W^-\\
    A^0 &\rightarrow Zh, t\bar{t}, b\bar{b}\\
    H^\pm &\rightarrow tb, W^\pm h
\end{align}
and the main production modes are
\begin{align}
    &gg, VV \rightarrow h', h''\\
    &gg \rightarrow A^0\\
    &b\bar{b} \rightarrow A^0\\
    &g\bar{b} \rightarrow t H^+.
\end{align}

The $13$~TeV LHC searches for heavy BSM Higgs bosons are~\cite{Aaboud:2017gsl,Aaboud:2017rel,ATLAS:2016qmt,TheATLAScollaboration:2016ibb,Aaboud:2017cxo,ATLAS:2016qiq,CMS:2017vpy,CMS:2016ncz}, and target all the decay channels shown in~\eqref{eq:decay} except $h'/A\rightarrow t\bar{t}$ and $H^\pm \rightarrow W^\pm h$, which is expected to be largely subdominant to $H^\pm \rightarrow tb$. In what follows, for each parameter point of our model, we compute the branching ratios corresponding to the decay modes shown in~\eqref{eq:decay} using the formulae given in~\cite{Djouadi:2005gj}. We take the production cross sections from the LHC Higgs cross section working group and rescale them by the following $\kappa$ factors
\begin{align}
    \kappa_{ggh'} &= \frac{(c_1c_2c_3 - s_1s_3)^2}{\tan^2\beta}\\
    \kappa_{ggA^0} &= \frac{1}{\tan^2\beta}\\
    \kappa_{bbA^0} &= \tan^2\beta\\
    \kappa_{VVh'} &= [-s_3 \cos(\beta - \alpha_1) + c_2 c_3 \sin(\beta - \alpha_1)]^2\\
    \kappa_{gbH^\pm} &= \left[-\frac{1}{\tan\beta} + \frac{m_b}{m_t}\tan\beta\right]^2
\end{align}
where the mixing parameters $\alpha_1$, $c_{123}$, and $s_{123}$ are defined in~\eqref{eq:rotmatrix}. For each channel, we finally compare the values of $\sigma \times \mathrm{Br}$ to the corresponding LHC limits.

\section{Parameter Space and Results}
\label{sec:parameter} 

We now scan over the parameter space of our multiple Higgs model to determine how much the fine-tuning can be lowered without breaking perturbativity or being at odds with the LHC results. Our model involves ten parameters: seven quartic couplings $\lambda_i$, the mixing angle $\beta$ between the vevs of the Higgs doublets, the trilinear coupling $A_h$, and the vev $u$ of the singlet $\Phi$. After requiring $m_h= 125$~GeV, the parameter space is then nine-dimensional.  Such a large parameter space is particularly difficult to explore. We would therefore like to stress that our final result will be conservative, as narrow regions with low fine-tuning might have been overlooked.

    In what follows, we choose a cutoff scale $\Lambda = 5$~TeV and perform a uniform random scan over the following parameters
    \begin{align}
    	u &\in [0, 5]~\mathrm{TeV}\\
	\lambda_i' &\in [-2, 2]\\
	\beta &\in \left[0, \frac{\pi}{2}\right]
    \end{align}
    and fix $A_h$ by setting the lightest Higgs mass to be $125$~GeV.  We emphasize choosing a common cutoff scale $\Lambda$ is very conservative.  The gauge boson, scalar and top sectors could have different cutoffs which would allow a much larger number of models that meet our scan criterion.  The $\lambda_i'$ couplings are linear combinations of the $\lambda_i$ couplings, and are defined in Appendix~\ref{sec:RGE}. These combinations are the ones who enter the RGEs and are therefore more relevant to scan over from a perturbativity point of view. 
    We scan over $10^9$ points and select the models verifying the vacuum stability and perturbativity constraints discussed in section~\ref{sec:model} and for which the couplings of the $125$~GeV Higgs to the gauge bosons are within $10$\% of the corresponding SM values. Additionally, in order to ensure that our results will not be influenced by the physics at the cutoff scale we consider only models where the new particles have masses below $1$~TeV. Finally, we select all points for which the fine-tuning factor $D$ is less than $100$. These points will be represented in blue in the figures shown in this section. For each of these points and for the different BSM Higgs bosons, we compute the cross-section times branching ratio for each of the production and decay channels listed in section~\ref{sec:lhc} and compare it to the results from the ATLAS searches~\cite{Aaboud:2017gsl,Aaboud:2017rel,ATLAS:2016qmt,TheATLAScollaboration:2016ibb,Aaboud:2017cxo,ATLAS:2016qiq}. We consider that a parameter point is not excluded at the LHC if $h'$, $A^0$, and $H^\pm$ are all heavier than $250$~GeV and, for each detection channel, the ratio of the $\sigma \times \mathrm{Br}$ over the $95$\% confidence limit found by ATLAS is less than $1$. In order to account for the important fluctuations of the ATLAS exclusion bounds at low Higgs masses as well as estimate the reach of the future LHC searches, we also define a ``safe'' region where the ratio of the $\sigma \times \mathrm{Br}$ over the $95$\% ATLAS confidence limit for each detection channel is less than $0.1$. 
\begin{figure}[t]
	\centering
	\includegraphics[width=\linewidth]{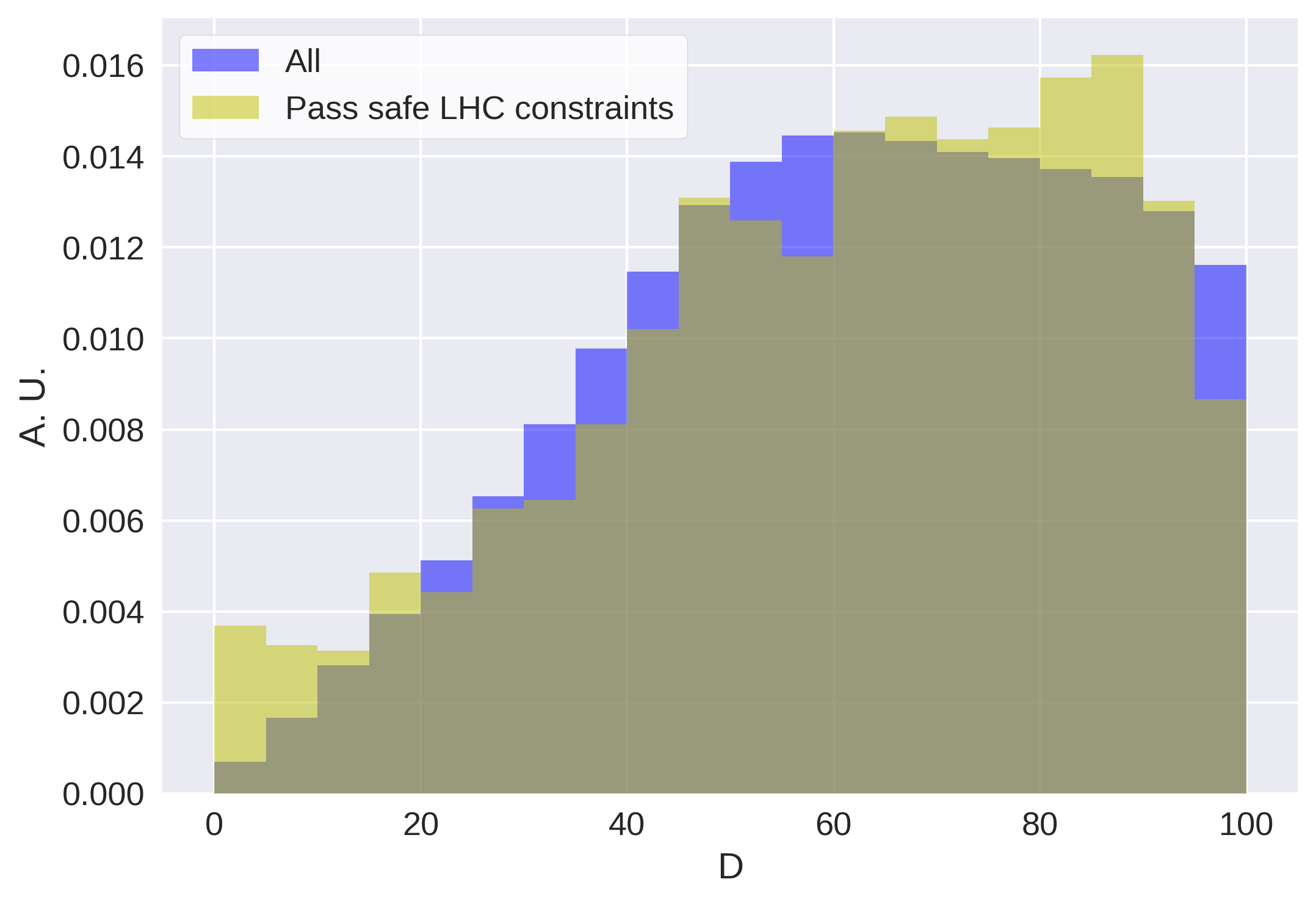}
	\caption{\label{fig:FT} Fine-tuning factor $D$ for all the points with $g_V > 0.9$ and $D < 100$ (blue) and for the subset of these points that satisfy the safe LHC constraints  defined in the main text (yellow).  The vertical axis is in arbitrary units with different scales for the blue and for the yellow.}
    \end{figure}

    % \begin{figure}
	% \centering
	% \includegraphics[width=\linewidth]{Plots/mh2_Ah.png}
	% \includegraphics[width=\linewidth]{Plots/mh3_Ah.png}
	% \caption{\label{fig:FT} $m_{h'}$ versus $A_h$ (top) and $m_{h''}$ versus $A_h$ (bottom) for all the points with $g_V > 0.9$ and $D < 100$ (blue dots), the points that satisfy the LHC constraints (red triangles), and the subset of these points that verify $D < 25$ (yellow squares).}
    % \end{figure}

    % \begin{figure}
	% \centering
	% \includegraphics[width=\linewidth]{Plots/FT_vs_mH2.png}
	% \caption{\label{fig:FTmH2} Fine-tuning factor $D$ for all the points with $g_V > 0.9$ and $D < 100$ (blue) and for the subset of these points that satisfy the LHC constraints (red).}
    % \end{figure}

    \begin{figure}[t]
	\centering
	\includegraphics[width=\linewidth]{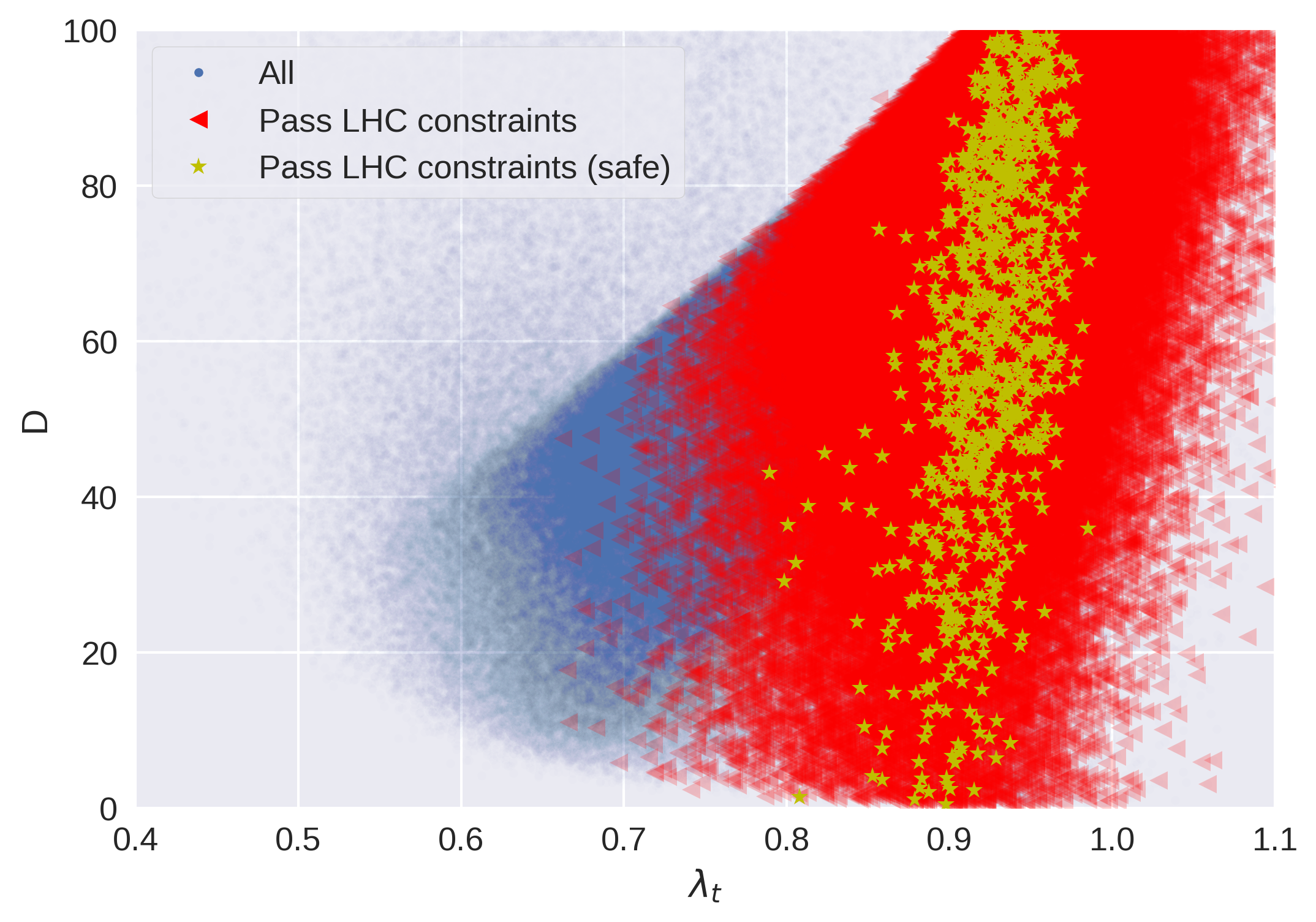}
	\caption{\label{fig:FTtop} Fine-tuning $D$ versus $\lambda_t$ for all the points with $g_V > 0.9$ and $D < 100$ (blue dots), the points that satisfy both $D < 20$ and the LHC constraints (red triangles), and the subset of these points that satisfy the ``safe'' LHC constraints (yellow stars).  Both LHC constraints are defined in Section~V.}
    \end{figure}

    \begin{figure}[t]
	\centering
	\includegraphics[width=0.95\linewidth]{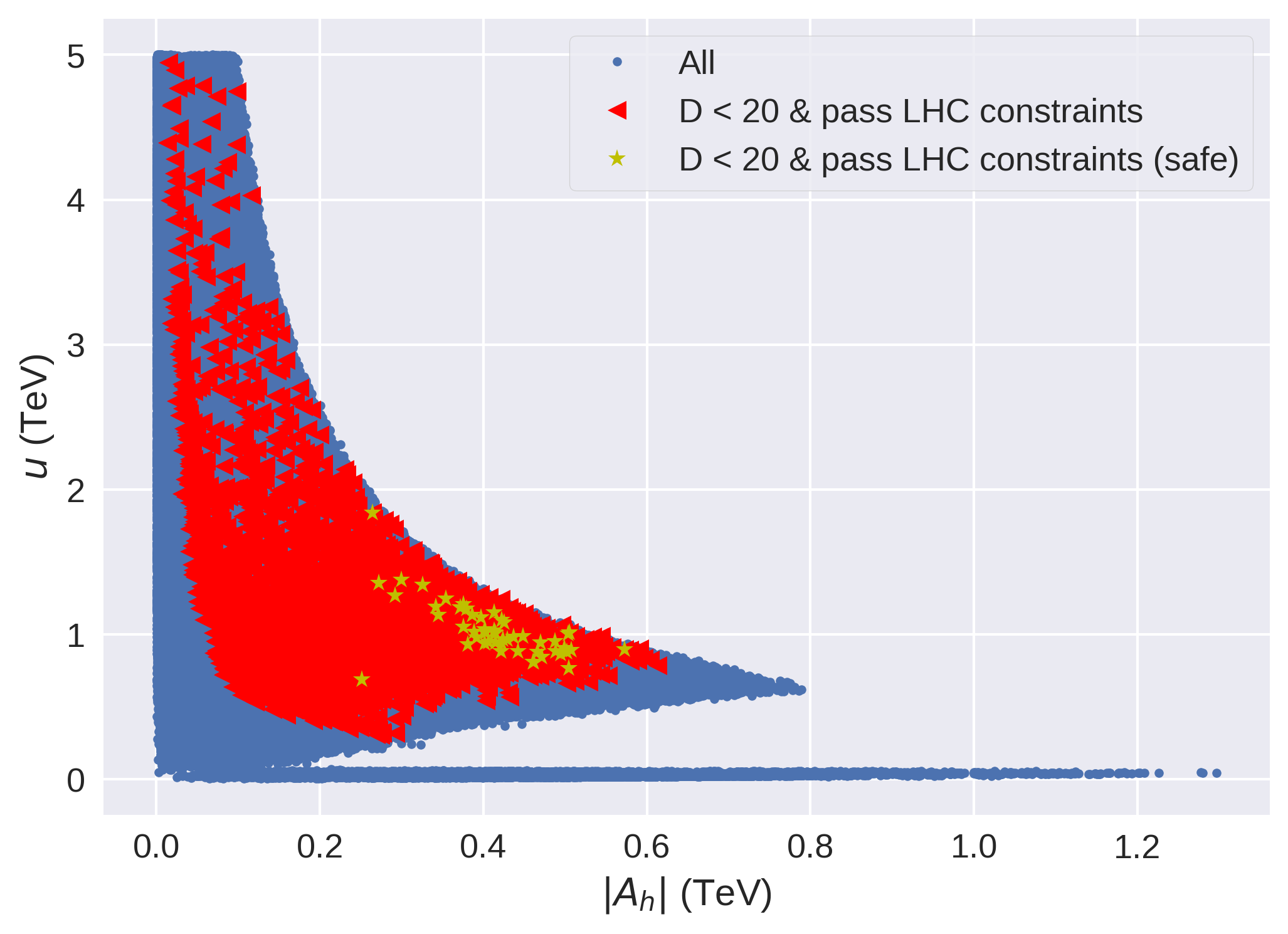}
	\caption{\label{fig:uAh} $u$ versus $A_h$ for all the points with $g_V > 0.9$ and $D < 100$ (blue dots), the points that satisfy both $D < 20$ and the LHC constraints (red triangles), and the subset of these points that satisfy the ``safe'' LHC constraints (yellow stars).}
    \end{figure}

    \begin{figure}[t]
	\centering
	\includegraphics[width=0.95\linewidth]{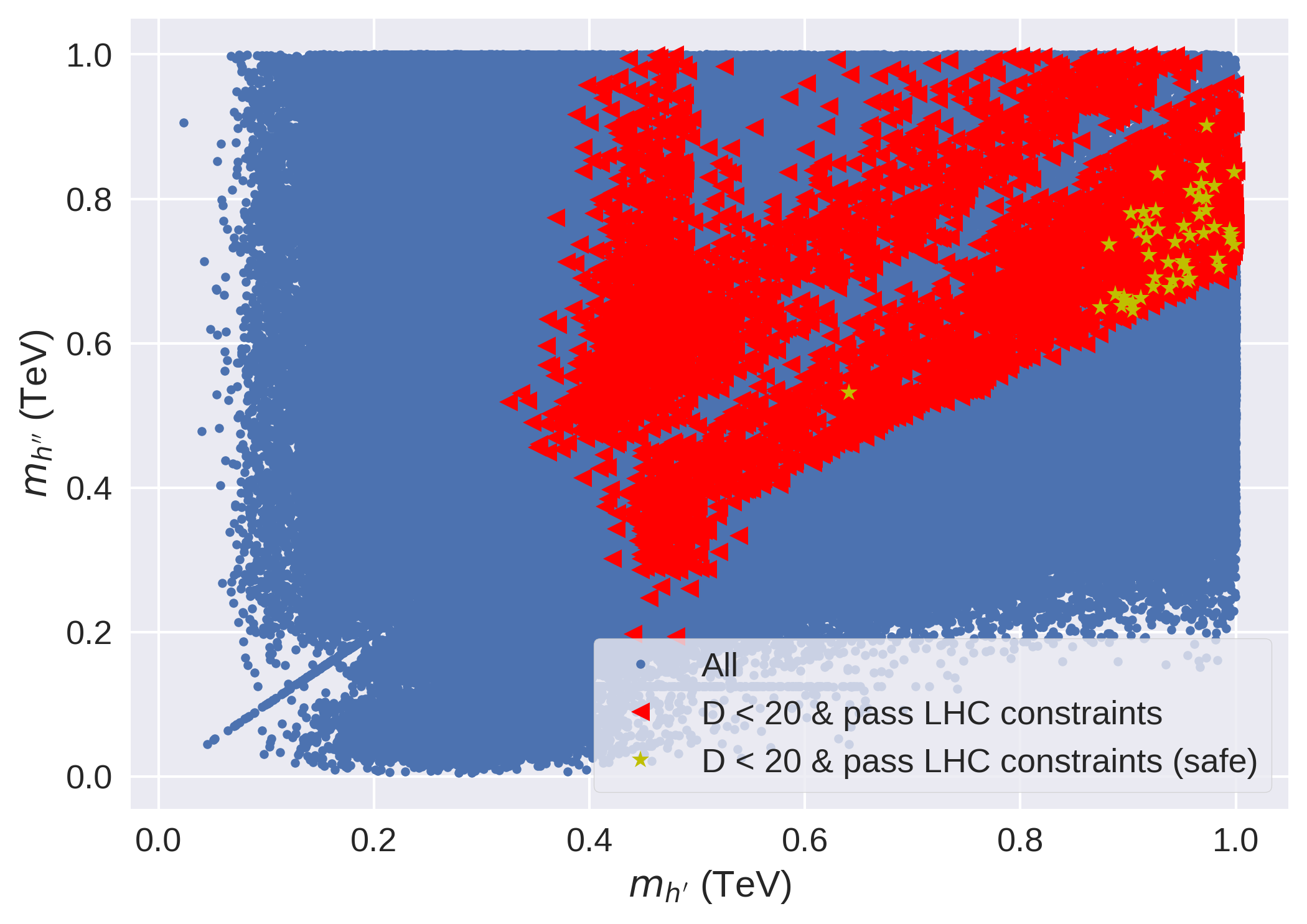}
	\caption{\label{fig:m3m2}  $m_{h''}$ versus $m_{h'}$ for all the points with $g_V > 0.9$ and $D < 100$ (blue dots), the points that satisfy the LHC constraints (red triangles), and the subset of these points that pass the ``safe'' LHC constraints (yellow stars).}
    \end{figure}

    \begin{figure}[t]
	\centering
	\includegraphics[width=0.95\linewidth]{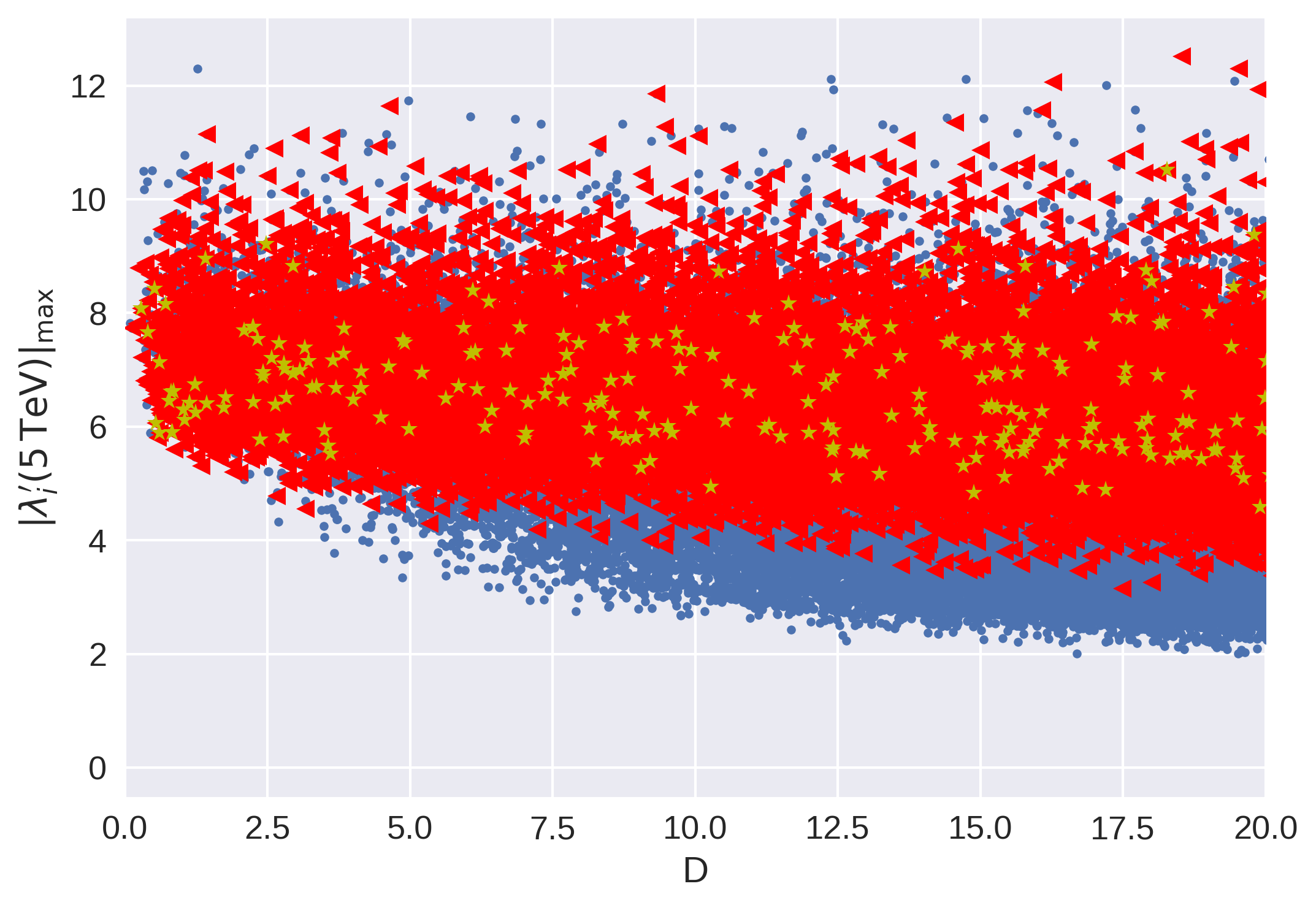}
	\caption{\label{fig:lft}  Maximal value of the couplings $\lambda_i'$ defined in Appendix~\ref{sec:RGE} at 5~TeV, $|\lambda_i'(5~TeV)|_\mathrm{max}$, versus fine-tuning for all the points with $g_V > 0.9$ and $D < 100$ (blue dots), the points that satisfy the LHC constraints (red triangles), and the subset of these points that pass the ``safe'' LHC constraints (yellow stars).}
    \end{figure}
    
 Figure~\ref{fig:FT} shows the normalized distributions of the fine-tuning factors $D$ for all the points with $g_V > 0.9$ and $D < 100$ (blue) as well as the subset of these points that satisfy the LHC constraints defined above (yellow). Whether the LHC constraints are taken into account or not, the fine-tuning factor $D$ can easily reach values smaller than $10$ or even $1$. Thus, lowering the top quark coupling while in the same time exploiting the partial cancellation of the top and scalar one-loop contributions to the Higgs mass could potentially be an way to reduce the fine-tuning in the SM without introducing too much complexity. 

    In order to understand the interplay between the suppression of the top yukawa coupling and the  cancellation from scalar couplings, we show the fine-tuning factor $D$ as a function of $\lambda_t$ in figure~\ref{fig:FTtop}. When the LHC constraints are not introduced, this figure shows two distinct low fine-tuning regions: one region where the top Yukawa coupling is reduced to values as low as $0.55$ for $D < 20$, and one where the top coupling to the SM Higgs is unsuppressed and the reduction of the fine-tuning is entirely due to the other scalars. The latter region  involves new Higgses that are typically heavy due to a large quartic coupling, and  therefore outside the reach of the LHC. Conversely, most of the first region has already been probed by the LHC. This result is due to the fact that such low values of the fine-tuning require the top coupling of the $125$~GeV Higgs $\lambda_t$ to be significantly reduced. In section~\ref{sec:gauge}, we already argued that $\tan\beta$ cannot be too small in order for the top Yukawa coupling of the $h_2$ to remain perturbative, which in turn requires models with suppressed $\lambda_t$ to be far from the alignment limit. This numerical study then shows that these two conflicting constraints prevent models with reduced $\lambda_t$ to have BSM Higgses heavier than a few hundreds of GeV. In fact, our results show that the only way to obtain suppressed top Yukawa couplings is to have large mixings between the scalars, which can happen only in the low mass regime. Although a few points with $D\lesssim 10$ still survive the current constraints, especially for $\lambda_t \gtrsim 0.7$, they are expected to be probed by the next LHC runs. The hypothesis that the fine-tuning of the $125$~GeV Higgs mass is reduced by suppressing the top quark Yukawa coupling should therefore be fully tested in the near future.

    Figures~\ref{fig:uAh} and \ref{fig:m3m2} show the mass scales corresponding to the regions of the parameter space with the lowest fine-tuning, in the $(u, A_h)$ plane and in the $(m_{h'}, m_{h''})$ plane. Although there are narrow regions with either $u \ll |A_h|$ or $m_{h''}\gg m_{h'}$, the low fine-tuning points that pass the safe LHC constraints generally have $|A_h|$ and $u$ being of the same order of magnitude. In most of the parameter space verifying these constraints, we also note that $h''$ is lighter than $h'$. Although most of the fully degenerate limit is already excluded, the points that survive the current LHC constraints generally have $m_{h'}\sim m_{h''}$.

    Finally, figure~\ref{fig:lft} shows the maximal value of the $\lambda_i'$ couplings defined in Appendix~\ref{sec:RGE} at the cutoff scale $\Lambda = 5$~TeV as a function of the fine-tuning $D$. Although, in principle, these couplings could reach values down to $2$ for $D < 20$, once the LHC constraints are introduced, these couplings have to be larger than $4$. This result is due to the fact that these LHC constraints disfavor the regions of parameter space with low top Yukawa $\lambda_t$ and large quartic couplings are therefore required in order to cancel the unsuppressed top loop contribution to the Higgs mass, as shown in equation~\leqn{eq:diverge}. Reducing the total fine-tuning in our model therefore implies the existence of a strongly coupled Higgs sector at a few TeV.
    
\section{Beyond}

We have discussed how the Higgs mass parameter is
calculable from the bare mass and radiative corrections, as shown in equation~\leqn{eq:diverge}.  In Section~\ref{sec:idea} we have posed the question of whether the bare mass and the one-loop radiative corrections need to be fine-tuned.  %  
%It would be interesting to know whether there needs to be finely tuned cancellation between them.
This is not possible in the minimal standard model unless the cutoff scale is below a TeV--a scale the LHC has substantially explored.  %In considering extensions of the Standard Model, so far we have focused on the radiative corrections, as these are affected by physics at LHC accessible energies.  
In the previous sections, we focused on a minimal model that illustrates aspects of the first and third Veltman conditions in Section~\ref{sec:idea}.  We now briefly discuss the second condition.  %The Higgs mass parameter is calculable from a combination of cutoff scale physics (a bare mass) and radiative corrections, it would be interesting to know whether there needs to be finely tuned cancellation between them. 
We briefly note that theories with softly broken shift symmetries can predict light scalars \cite{Georgi:1975tz,ArkaniHamed:2001nc}.  %
%
%However if the radiative corrections to a small Higgs mass are small, then the bare mass also needs to be small. 
%
Extensions of this idea developed into little Higgs model building \cite{ArkaniHamed:2001nc,Schmaltz:2002wx,Katz:2005au}.  Here %  have generally introduced 
top partners were introduced to cancel the quadratic sensitivity to the cutoff scale from radiative corrections due to the top coupling.  Given the second Veltman condition, it is conceivable that the cancellations just occur between the top and additional scalars for no obvious symmetry reason, or that the observed scalar does not have such a large top coupling as to require top partners.  We consider this possibility in future work.  

We also note that there is lattice gauge theory evidence that some strongly coupled theories  have scalars which are much lighter than the strong coupling scale for a dynamical reason  which is not related to a symmetry~\cite{Hasenfratz:2017lne}.  

\section{Conclusions}
 We have considered the cutoff sensitivity of an extension of the minimal standard model with additional  particles which only carry electroweak charges, and argue that they should be scalars. Our goals are more modest that those of theorists who achieve   cancellations in the Higgs mass from symmetry considerations, but  we do not have to pay the price of introducing new colored top partners or a new strong group, which therefore makes it  easier to understand how the new physics has escaped the LHC searches. We do find that there are values of the parameters for which the cancellation between the bare mass and the radiative corrections is not severe and the LHC constraints are satisfied, including points where the top Yukawa coupling to the 125 GeV higgs is reduced and the cutoff scale for new physics is 5 TeV.  Our results emphasize the importance of precision Higgs measurements, particularly direct measurements of the top-Higgs coupling, as we find there is room in this model for significant deviation from 1, and this value impacts the degree of fine-tuning and the expectation for the scale of new physics. Searches for additional Higgs bosons are also important  in order to understand whether the naturalness paradigm that has played such a big role in theoretical physics could be realized in nature. 

\section*{Acknowledgements}
SEH acknowledges support by the NWO Vidi grant ``Self-interacting asymmetric dark matter''. AN is supported in part by the Kenneth Young Memorial Endowed Chair and in part by the DOE under grant DE-SC0011637.  DW is supported in part by a Burke faculty fellowship.\\

\appendix
\section{SCALAR MASS EIGENSTATES}
\label{sec:eig}
In the limit where $\lambda_i \ll 1$ and $u, v_1, v_2 \ll |A_h|$, the same neutral mass eigenstates are
\begin{widetext}
\begin{eqnarray}
m_1^2 &=& 2 A_h u + \frac{2}{v_1 v_2} \Bigl(\left(\lambda_1 v_1^2+\lambda_4 (v_1+v_2)^2+\lambda_2 v_2^2\right) v_1 v_2- \left(\left(\lambda_1+2 \lambda_4-\lambda_6\right)v_1  + \left(\lambda_2+2 \lambda_4-\lambda_7\right)v_2\right) u^2 (v_1+v_2)\Bigr) \hspace{1cm} \\
&-& \frac{1}{2 A_h u} \Bigl(\bigl( (\lambda_1+\lambda_4) (3 \lambda_1-2 \lambda_2+\lambda_4-4 \lambda_6+4 \lambda_7)\,v_1^2+ (\lambda_2+\lambda_4) (-2 \lambda_1+3 \lambda_2+\lambda_4+4 \lambda_6-4 \lambda_7)\,v_2^2\bigr)u^2 \nonumber  \\
&+& 4 (\lambda_1+\lambda_4) (\lambda_2+\lambda_4)\,v_1^2 v_2^2 \Bigr)  + \ldots \nonumber \\
m_2^2 &=& -2 A_h u + \frac{2}{v_1 v_2} \Bigl( \bigl( (\lambda_1+2 \lambda_4-\lambda_6)v_1- (\lambda_2 + 2 \lambda_4 -\lambda_7)v_2 \bigr)u^2 (v_1-v_2) \\
&+& \left( (\lambda_1+\lambda_4)v_1^2 -2 \lambda_4 v_1 v_2+ (\lambda_2+\lambda_4)v_2^2 \right)v_1 v_2\Bigr) + \frac{1}{2 A_h u} \Bigl( (\lambda_1+\lambda_4) (3 \lambda_1-2 \lambda_2+\lambda_4-4 \lambda_6+4 \lambda_7)v_1^2 \nonumber \\
&+&(\lambda_2+\lambda_4) (-2 \lambda_1+3 \lambda_2+\lambda_4+4 \lambda_6-4 \lambda_7)v_2^2 \Bigr) u^2 +4 (\lambda_1+\lambda_4) (\lambda_2+\lambda_4) v_1^2 v_2^2 + \ldots \nonumber \\
m_3^2 &=& 4 u^2 (\lambda_1+\lambda_2+\lambda_3+4 \lambda_4-\lambda_6-\lambda_7) - A_h  + \left(  \frac{v_1 v_2}{u} + \frac{u v_1}{v_2} + \frac{u v_2}{v_1} \right) \ldots \nonumber
\end{eqnarray}
\end{widetext}
\null\clearpage
\section{COLEMAN-WEINBERG POTENTIAL}
\label{sec:CWpotential}

The scalar mass matrix, derived from equation~\leqn{eq:scalarpotential}, has the form $v^\dagger M^2_s\, v$ with $v = \left(h_1, h_2, \phi \right)$.  Throughout this work, we require the scalar potential couplings to be $\lambda_i < 1$.  To understand, the form of the effective potential, first consider the limit where $\lambda_i  = 0$.  $M^2_s$ is simply
\begin{equation}
M^2_s = {A_h \over 3}\begin{pmatrix} -3\,u\,v_2/v_1 & \phi & h_2  \\ \phi & -3\,u\,v_1/v_2 & h_1 \\ h_2^\dagger & h_1^\dagger & -3\,v_1\,v_2/u \end{pmatrix} 
\end{equation}
In this limit, $M_s$ satisfies
\begin{align}
{\partial \over \partial \theta}\mathrm{tr}M_s^2 = 0 % && {\partial \over \partial \theta}\mathrm{tr}\log{\left(M_s^2/\Lambda^2\right)} \,.
\end{align}
where $\theta = \{h_1, h_2, h_1^\dagger, h_2^\dagger,\phi \}$.  This ensures that the effective potential has no quadratic diverging terms proportional to $A_h^2$.  Moreover, the largest logarithmic terms are proportional to $A_h^2$.  

We find that the quadratically diverging terms in the effective potential are 
\begin{eqnarray}\label{eq:quadraticCW}
V_\mathrm{quadratic} &=& %{\Lambda^2 \over 32\,\pi^2}  \mathrm{tr}\biggl[M_\mathrm{s}^\dagger M_\mathrm{s}  \biggr]%-  {3 \Lambda^2 \over 8\pi^2} M_\mathrm{top}^\dagger M_\mathrm{top}
% \nonumber \\
%&=& 
 {\Lambda^2 \over 32\,\pi^2} \left( \lambda_1+2 \lambda_4+\frac{1}{2}\lambda_5 +\frac{1}{2}\lambda_6\right) h_1^\dagger h_1 \\
&+&{\Lambda^2 \over 32\,\pi^2} \left( \lambda_2+2 \lambda_4+\frac{1}{2}\lambda_5+\frac{1}{2}\lambda_7 \right) h_2^\dagger h_2 \nonumber\\ 
&+& {\Lambda^2 \over 32\,\pi^2} \left(\lambda_3+{\lambda_6 \over 2}+\frac{\lambda_7}{2}\right) \phi ^2 - {3\,\lambda^2\,\Lambda^2 \over 8\,\pi^2} h_2^\dagger h_2  \nonumber
%&-&  {3 \Lambda^2 \over 8\pi^2}  \nonumber
\end{eqnarray}
where $\Lambda$ is the cutoff.  We have omitted the contributions to the cosmological constant.  The term proportional to $\lambda^2$ is generated by equation~\leqn{eq:lambda}.  The largest logarithmically divergent terms are
\begin{eqnarray}
V_\mathrm{logarithm} &=& {1 \over 64\pi^2}\sum_{i = 1}^3  M_i^4 \log\left[{M_i^2 \over \Lambda^2} \right]  \\
&-& {3\,\lambda^4 \over 16\pi^2} \left( h_2^\dagger h_2 \right)^2\log\left[{\lambda^2 h_2^\dagger h_2 \over \Lambda^2} \right] \nonumber 
%&+& \phi^2 \,\log\left( {A_h^2  u^2  \over  \Lambda^4}\right) \biggr) \log\left( {A_h\, v_1\, v_2  \over u \,\Lambda^2}\right) + \ldots  \nonumber \\
%{1 \over 64\,\pi^2} \left(M_\mathrm{s}^\dagger M_\mathrm{s} \right)^2 \log\left( {M_\mathrm{s}^\dagger M_\mathrm{s} \over \Lambda^2 } \right) \\
%&-& {3\,\lambda^4 \over 16\,\pi^2} \,h_2^4 \, \log\left(\lambda^2\, h_2^2\, /\Lambda^2 \right) 
\end{eqnarray}
where $M_i^2$ is defined by
\begin{widetext}
\begin{eqnarray}
M_1^2 &=& -\frac{A_h u}{2 \,v_1 v_2}\left(v_1^2+v_2^2\right) -\left(\lambda_1+2 \lambda_4\right) v_1^2- \left(\lambda_2+2 \lambda_4\right) v_2^2 + \frac{A_h u}{18 \,v_1 v_2}\left(h_1 h_1^\dagger+h_1 h_2^\dagger+h_2^\dagger h_1+h_2 h_2^\dagger\right) \\
&+&\frac{A_h }{3}\phi -\frac{\lambda_5}{4} \left(h_1^\dagger h_2+ h_2^\dagger h_1\right)+\frac{1}{4}  \left(2 \lambda_1+4 \lambda_4+\lambda_5\right) h_1^\dagger h_1+\frac{1}{4} \left(2 \lambda_2+4 \lambda_4+\lambda_5\right)h_2^\dagger h_2+\frac{1}{4}  (\lambda_6+\lambda_7)\,\phi ^2 \nonumber  \\
M_2^2 &=& -\frac{A_h u }{2\, v_1 v_2}\left(v_1^2+v_2^2\right) - \left(\lambda_1+2 \lambda_4\right) v_1^2- \left(\lambda_2+2 \lambda_4\right)v_2^2+ \frac{A_h u}{18\, v_1 v_2}\left(h_1 h_1^\dagger-h_1 h_2^\dagger-h_2^\dagger h_1+h_2 h_2^\dagger\right)\\
&-&\frac{A_h }{3}\phi  + \frac{\lambda_5}{4} \left(h_2^\dagger h_1+h_1^\dagger h_2\right)+\frac{1}{4}  \left(2 \lambda_1+4 \lambda_4+\lambda_5\right)\,h_1^\dagger h_1+\frac{1}{4} (2 \lambda_2+4 \lambda_4+\lambda_5)\,h_2^\dagger h_2  +\frac{1}{4} (\lambda_6+\lambda_7) \,\phi^2 \nonumber \\
M_3^2 &=& -\frac{A_h v_1 v_2}{u}-\frac{A_h u}{9\, v_1 v_2}\left(h_1 h_1^\dagger+h_2 h_2^\dagger\right)+ \frac{\lambda_6}{2}  h_1^\dagger h_1+\frac{\lambda_7}{2}  h_2^\dagger h_2+\lambda_3 \,\phi ^2-\left(\lambda_6 \,v_1^2+\lambda_7 \,v_2^2\right) \, .
\end{eqnarray}
\end{widetext}
Here we have assumed that  $u \gg \Lambda$, $\lambda_i < 1$.

\section{RENORMALIZATION GROUP EQUATIONS}
\label{sec:RGE}
    Before canceling all the tadpole terms, the Higgs potential that we are using could be rewritten as
    \begin{widetext}
	\begin{eqnarray}
	\nonumber
    V =& \frac{\lambda_1'}{2} (h_1^\dagger h_1)^2 + \frac{\lambda_2'}{2} (h_2^\dagger h_2)^2 + \frac{\lambda_3'}{24} \phi^4 + \lambda_4' (h_1^\dagger h_1)(h_2^\dagger h_2) + \lambda_5' (h_1^\dagger h_2)(h_2^\dagger h_1) + \frac{\lambda_6'}{2} \phi^2 (h_1^\dagger h_1) \\
    &+ \frac{\lambda_7'}{2} \phi^2 (h_2^\dagger h_2) + A_h(\phi h_1h_2^\dagger + \phi h_2 h_1^\dagger). 
	\end{eqnarray}
    \end{widetext}
    \null\clearpage
    \null The primed couplings are related to the original couplings by
    \begin{align}
	\lambda_1' &= 2(\lambda_1 + \lambda_4)\\
	\lambda_2' &= 2(\lambda_2 + \lambda_4)\\
	\lambda_3' &= 24 \lambda_3\\
	\lambda_4' &= 2\lambda_4 + \lambda_5\\
	\lambda_5' &= -\lambda_5\\
	\lambda_6' &= 2\lambda_6\\
	\lambda_7' &= 2\lambda_7
    \end{align}
    and should remain of order one. The beta functions for the 2HDM are well-known~\cite{Alanne:2016wtx}. We derive the contributions from the $\phi$ field as well as the RG equation for $A_h$ from the Coleman-Weinberg potential. We checked our values for the contributions from the quartic couplings involving $\phi$ against~\cite{Alanne:2016wtx}. Neglecting the weak couplings, we obtain
    \begin{widetext}
    \begin{eqnarray}
    	16\pi^2 \frac{d\lambda_1'}{dt} &=& 12 \lambda_1^2 + 4\lambda_4'^2 + 4 \lambda_4'\lambda_5' + 2\lambda_5'^2 + \lambda_6'^2\\
    	16\pi^2 \frac{d\lambda_2'}{dt} &=& 12 \lambda_2^2 + 4\lambda_4'^2 + 4 \lambda_4'\lambda_5' + 2\lambda_5'^2 + \lambda_7'^2 + 12 \lambda_t^2\lambda_2' - 12\lambda_t^4\\
    	16\pi^2 \frac{d\lambda_3'}{dt} &=& 6\lambda_3'^2 + 12(\lambda_6'^2 + \lambda_7'^2)\\
    	16\pi^2 \frac{d\lambda_4'}{dt} &=& (\lambda_1' + \lambda_2')(6\lambda_4' + 2\lambda_5') + 4\lambda_4'^2 + 2\lambda_5'^2 + \lambda_6'\lambda_7'+ 6 \lambda_t^2\lambda_4'\\
    	16\pi^2 \frac{d\lambda_5'}{dt} &=& 2(\lambda_1' + \lambda_2') \lambda_5' + 8\lambda_4'\lambda_5' + 4\lambda_5'^2 + 6\lambda_t^2\lambda_5'\\
    	16\pi^2 \frac{d\lambda_6'}{dt} &=& 4 \lambda_6'^2 + 6\lambda_1'\lambda_6' +  \lambda_3'\lambda_6' + 4\lambda_4'\lambda_7' + 2\lambda_5'\lambda_7'\\
    	16\pi^2 \frac{d\lambda_7'}{dt} &=& 4 \lambda_7'^2 + 6\lambda_2'\lambda_7' +  \lambda_3'\lambda_7' + 4\lambda_4'\lambda_6' + 2\lambda_5'\lambda_6'\\
    	16\pi^2 \frac{dA_h}{dt} &=& 8 A_h(\lambda_4' + 2\lambda_5' + \lambda_6' + \lambda_7').
    \end{eqnarray}
    Here, $t = \log\mu$ (with $\mu$ the renormalization scale). Now, the RG equations for the Yukawa and gauge couplings (also including the strong coupling $g_s$) are also well-known and are~\cite{Peskin:1997ez}
    \begin{align}
    	16\pi^2 \frac{d\lambda_t}{dt} &= \frac{9}{2}\lambda_t^3 - 8 g_s^2\lambda_t\\
	16\pi^2 \frac{dg_s}{dt} &= -7 g_s^3\\
    \end{align}
\end{widetext}
\bibliography{biblo}

\end{document}